\documentclass[preprint,12pt,authoryear]{elsarticle}
\usepackage{amssymb}
\usepackage{times}
\usepackage{amsmath}

\journal{New Astronomy}

\begin{document}

\begin{frontmatter}

\title{Wolf-Rayet stars, black holes and the first detected gravitational wave source}

\author[1]{A. I. Bogomazov}

\author[1]{A. M. Cherepashchuk}

\author[1]{V. M. Lipunov}

\author[2]{A. V. Tutukov}

\address[1]{M. V. Lomonosov Moscow State University, P. K. Sternberg Astronomical Institute, 13, Universitetskij prospect, Moscow, 119991, Russia}

\address[2]{Institute of astronomy, Russian Academy of Sciences, 48, Pyatnitskaya ulitsa, Moscow, 119017, Russia}

\begin{abstract}
The recently discovered burst of gravitational waves GW150914 provides a good new chance to verify the current view on the evolution of close binary stars. Modern population synthesis codes help to study this evolution from two main sequence stars up to the formation of two final remnant degenerate dwarfs, neutron stars or black holes \citep{massevich1988}. To study the evolution of the GW150914 predecessor we use the ``Scenario Machine'' code presented by \citet{lipunov1996}. The scenario modelling conducted in this study allowed to describe the evolution of systems for which the final stage is a massive BH+BH merger. We find that the initial mass of the primary component can be $100\div 140 M_{\odot}$ and the initial separation of the components can be $50\div 350 R_{\odot}$. Our calculations show the plausibility of modern evolutionary scenarios for binary stars and the population synthesis modelling based on it.
\end{abstract}


\end{frontmatter}

\section{Introduction}

The LIGO discovery of the first gravitational wave burst GW150914 (Abbott et al., 2016a) starts a new era in physical and astrophysical studies (Abbott et al., 2016b,c). The detector observed a merger of two black holes with masses $36^{+5}_{-4} M_{\odot}$ and $(29\pm 4) M_{\odot}$, the amount of energy release in this merger in the form of gravitational waves was $(3\pm 0.5) M_{\odot} c^2$, the mass of the remnant single black hole was $(62\pm 4) M_{\odot}$, and its spin was $0.67^{+0.05}_{-0.07}$. This paper aims at a study of the possible evolution of a close massive binary star that leads to a merger of two black holes with parameters similar to those of GW150914.  For our calculations we use the ``Scenario Machine'', see the latest version of its detailed description by \citet{lipunov2009}. A huge number of papers were dedicated to investigations of the evolution of close binary stars and its implications to parameters of mergers and merging rates of relativistic stars, e. g. \citep{tutukov2002,belczynski2016,mink2016}.

\citet{lipunov1997a} studied merging rates of neutron stars (NS) with neutron stars, neutron stars with black holes (BH), and black holes with black holes (i.e. NS+NS, NS+BH, BH+BH) under different assumptions of the BH formation. The BH+BH and NS+NS merger rates in a galaxy like the Milky Way were found to be, respectively, $(2-5)\cdot 10^{-5}$ and $\sim 10^{-4}$ per year. A typical BH is formed with a mass 3-10 times the NS mass (which was assumed to be $1.4 M_{\odot}$), so the expected detection rate of BH+BH merging by LIGO was found to be 10-100 times higher than the NS+NS merger detection rate for a wide range of evolutionary parameters. Therefore \cite{lipunov1997a} concluded that the first LIGO event should be the BH+BH merger. \cite{belczynski2016} came to the same conclusion.

The discovery of gravitational waves by LIGO allows to trace an evolutionary scenario for massive binaries up to its final point and to estimate the possible range of evolutionary parameters. Here we study some of them for the most massive merging BHs in connection with GW150914. One of the crucial parameters of the evolution of binaries is the stellar wind mass loss. In the Section \ref{section-wind} we describe the mass loss rate by massive stars that can be progenitors of compact stars. In the Section \ref{black-wolf} we describe existing binary systems that consist of BHs and Wolf-Rayet (WR) stars that can be progenitors of massive merging binary BHs, also we mention the most massive close binaries with non-degenerate companions. In the Section \ref{scenario} we briefly describe the ``Scenario Machine'' program. In the Section \ref{results} the results of our study are presented. In the Section \ref{final} we make final conclusions and some discussions of the results of this paper.

\section{Stellar wind}

\label{section-wind}

Before 1990-ies the influence of mass loss in a form of stellar wind on the evolution of WR stars was highly overestimated. For example, \citet{langer1989a,langer1989b} published the following formula to connect mass loss of WR stars $\dot M_{WR}$ and their masses $M_{WR}$:

\begin{equation}
\dot M_{WR}=-(0.6-1.0)\cdot 10^{-7} \left(\frac{M_{WR}}{M_{\odot}}\right)^{2.5},
\label{langer}
\end{equation}

\noindent where the coefficient $0.6$ corresponds to WNE stars, and the value $1.0$ corresponds to WC and WO stars. Equation (\ref{langer}) leads to the so-called convergence effect: the mass of a WR star in the end of the evolution and the mass of its carbon-oxygen (CO) core does not exceed a few solar masses ($M_{CO}=2\div 4 M_{\odot}$) practically independently of its initial mass.

But how in this case are we able to understand the existence of black holes with masses in the range $10-15M_{\odot}$, that is a reliable observational fact \citep{cherepashchuk2013}? Moreover, non-LTE models of WR stellar winds by \citet{hillier1991} lead to unrealistically high mass loss rates: $\dot M> 10^{-5}M_{\odot}$ yr$^{-1}$, up to $10^{-4}M_{\odot}$ yr$^{-1}$. The effects of light scattering on electrons at so high $\dot M$ should produce observable wings in emission line profiles. A high $\dot M\sim 10^{-4}M_{\odot}$ yr$^{-1}$ also should lead to very deep blue shifted absorption components (like P Cyg) in the emission lines of WR stars. Such features are not observed in WR spectra. All these problems can possibly be resolved by the model of clumpy stellar winds of Wolf-Rayet stars. Most of existing data on mass loss rates of some tens of WR stars (and for O-B stars of I-II luminosity classes) are obtained on the basis of the analysis of their radio and infra-red (IR) thermal emission. \citet{cherepashchuk1990,cherepashchuk1991} noted that the presence of clumps in a stellar wind lead to overestimated values $\dot M$ for such stars, if the presence of clumps is ignored, because the thermal emission quadratically depends on the electron density. If the matter of the wind of the WR star is contained in numerous dense clumps, the intensity of the IR and radio emission grows in comparison with that for a uniform wind, so the real $\dot M$ value is overestimated.

Clumps in the stellar wind of a WR star were discovered by \citet{cherepashchuk1984} by analysis of atmospheric eclipses in the V444 Cyg binary system \\ (WN5+O6) in IR. They found that characteristic dimensions of the WN5 star's extended atmosphere are much greater in IR than in the optics, and they concluded that this wind was clumpy. Some years later \citet{moffat1988} obtained spectra of WR stars with very high signal to noise ratio ($\sim 300$) and found that the peaks of profiles of emission lines are variable: there are a lot of sharp emission components with the amplitude $\sim 1\%$ from the total line height that move along the lines. This fact directly proves the existence of clumps in WR stellar winds, and the clumps move out of these stars with an acceleration.

Photometric and polarization observations of WR stars in close binary systems gave more information about clumpy WR stellar winds. The realistic mass loss rate of a WR star estimated using the increase of the orbital period of the V444 Cyg eclipsing binary is $0.6\cdot 10^{-5} M_{\odot}$ yr$^{-1}$ \citep{cherepashchuk2013}. At the same time the value of $\dot M$ estimated using the analysis of IR and radio fluxes of this system is $\dot M_{WR}\approx 2.4\cdot 10^{-5} M_{\odot}$ yr$^{-1}$ \citep{prinja1990,prinja1991,howarth1992}. Observations of the linear polarization variability of some tens WR stars in close binaries \citep{stlouis1988} also lead to values of $\dot M$ for WR stars several times less in comparison with values found from their IR and radio fluxes.

\citet{cherepashchuk2001} calculated the final masses of WR stars and their carbon-oxygen (CO) cores under the assumption of clumpy WR winds \citep{cherepashchuk1990,cherepashchuk1991}. This allows to decrease the values of $\dot M_{WR}$ by a factor of $3\div 5$. \citet{cherepashchuk2001} used the following empirical formula (obtained from polarimetry observations of close binary WR+OB stars) to connect $\dot M_{WR}$ and $M_{WR}$:

\begin{equation}
\dot M_{WR}=KM^{\alpha}_{WR},
\label{kwr1}
\end{equation}

\noindent here $\alpha$ is in the range $1\div2$, and $\alpha=1$ is the more preferable value \citep{stlouis1988}. The decrease of the mass loss rate of WR stars by a factor of three and the low power in Equation (\ref{kwr1}) in comparison to Equation (\ref{langer}) allow to escape the convergence effect \citep{cherepashchuk2001}: the masses of CO cores of WR stars in the end of their evolution (they are the direct progenitors of relativistic objects) are in the wide range: $M_{CO}^{fin}=(1-2)M_{\odot}\div (20-40)M_{\odot}$. This interval of final masses of CO cores of WR stars includes the current observable interval of masses of neutron stars and black holes in X-ray binary systems: $M_{BH,NS}=1M_{\odot}\div 16M_{\odot}$.

In summary we can conclude that the standard $\dot M_{WR}$ derived from their IR and radio fluxes should be reduced by a factor of $3\div 5$, see e. g. \citep{hillier2003}. In calculations of modern extended atmospheres of WR stars in the non-LTE approximation the clumpy wind is defined arbitrarily: one defines a porosity parameter of clumps and an average density jump in them. Values of these parameters are obtained in analysis of line profiles in spectra of WR stars, see, e. g. \citep{hillier2003}. Stellar winds of massive hot OB and WR stars accelerated by radiation pressure in lines can be unstable to small perturbations of the wind density according to theoretical investigations, see, e. g. \citep{puls2003}. Also we have to mention that the WR wind can be non-symmetrical (which introduces an uncertainty in the mass-loss rates of a factor 2-3).

\section{Massive optical and degenerate binaries}

\label{black-wolf}

\citet{tutukov1973a,tutukov1973b} theoretically studied the possibility of the formation of WR stars in massive close binaries and suggested arguments in favour of the mass and angular momentum loss during the mass transfer before the formation of WR stars. The evolution of such binaries can include a stage with the close binary that consists of the WR star and the compact star. \citet{heuvel1972} considered Cen X-3 as an observational example of the binary system in the second mass exchange process that can preceed the formation of ``WR+compact star'' system, as was calculated \\ by \citet{heuvel1973}.

At the present time there are three known black holes in close binaries with Wolf-Rayet stars: Cyg X-3, IC 10 X-1, NGC 300 X-1, and probable fourth system CXOU J123030.3+413853 \citep{esposito2013}. In addition there are some systems that also are important to note for the aim of this paper: SS 433 (the only known super-Eddington accretor in the Milky Way at present time, this system can be a precursor of BH+WR binary), M33 X-7 (one of the most massive known black holes with a slightly evolved massive non-degenerate star), WR 20a, WR 22 (HD 92740), WR 21a, NGC 3603-A1, HDE 311884, and R 145 (LMC) that are close binaries with the most massive WR and O stars. Let us to discuss these binaries in some detail.

{\bf WR20a} was suggested to be a possible WR star by \citet{shara1991}, and found as a possible binary system by \citet{hucht2001}, because it has relatively weak emission lines in its spectrum that possibly belong to the secondary star. \citet{bonanos2004} estimated masses of the components from OGLE photometric observations and analysed spectral observations conducted by \citet{rauw2004}. As the result they obtained the masses of components to be $83\pm 5 M_{\odot}$ and $82 \pm 5 M_{\odot}$, and the orbital period to be 3.686 days. \citet{rauw2004} gave the following values of the system parameters: the orbital period is 3.675 days, lower limits of  masses of components are $70.7\pm 4 M_{\odot}$ and $68.8 \pm 3.8 M_{\odot}$ , and their spectral types are WN6ha or O3If, respectively.

{\bf WR 21a} is composed of a probable WNh star and an O star \citep{niemela2008}. The WNh component is the first WR star detected due to its bright X-ray emission \citep{caraveo1989,mereghetti1994}. The method to search for close binary systems among WR stars from X-ray observations of their colliding winds was proposed by \citet{cherepashchuk1976}. WR21a also is a source of non-thermal radio emission \citep{benaglia2005}. X-rays and radio rays indicate a
region of wind collision. WR 21a in projection lies close to Westerlund 2, so it could be ejected from this cluster. \citet{tramper2016} presented spectroscopic observations of WR21a. They obtained minimum masses of the components of $64.4\pm 4.8 M_{\odot}$ and $36.3\pm 1.7 M_{\odot}$, and derived spectral types as O3/WN5ha and O3V for primary and secondary stars correspodingly. Using the spectral type of the secondary as an indication for its mass, they estimated an orbital inclination of $i=58.8\pm 2.5^{\circ}$ and masses as $103.6\pm 10.2 M_{\odot}$ and $58.3\pm 3.7 M_{\odot}$ in agreement with the luminosity of the system.

{\bf WR 22} is a close binary system, the primary component is a WR star with mass $55.3 \pm 7.3 M_{\odot}$, the secondary component is a main sequence (MS) star with mass $20.6\pm 1.7 M_{\odot}$, the orbital period is $80.336 \pm 0.0013$ days \citep{rauw1996,schweickhardt1999}.

The abundances of stars like WR 20a, WR 22, and OB super giants were studied by \citet{bogomazov2008b} using the ``Scenario Machine''. They concluded that the WR20a system most likely is composed of a WR star with a MS star, or of two MS stars. The probabiltiy that the system consists of two WR stars is very low, but such systems should exist.

{\bf NGC 3603-A1} is an extremely luminous WN6+WN6 star \citep{drissen1995}, with luminosity in excess of $10^6 L_{\odot}$ \citep{koter1997,crowther1998} that shows radial velocity variations with a period of 3.772 days \\ \citep{moffat1984}, \citep{moffat1985}, \citep{moffat2004}. \citet{schnurr2008} estimated masses for the components of NGC 3603-A1 using the radial velocities for both components and the previously known inclination angle of the system as $116\pm 31 M_{\odot}$ for the primary and $89\pm 16 M_{\odot}$ for the secondary.

{\bf HDE 311884} is a massive binary star with WN6 and O5V components. The masses of the stars in the system are $\approx 51$ and $\approx 60$ masses of the Sun respectively \citep{hucht2001}.

\citet{schnurr2009} presented the results of a spectroscopic and polarimetric study of {\bf R145}. They combined radial velocity data from previous studies with previously unpublished polarimetric data. The orbital period of R145 is 158.8 days, the inclination angle $i$ was found to be $i = 38\pm 9^{\circ}$. They found minimum masses $M_{WR} \sin^3 i = 116\pm 33 M_{\odot}$ and $M_O \sin^3 i = 48\pm 20 M_{\odot}$ for the WR and the O component correspondingly. According to \citet{schnurr2009} the resulting absolute masses of the components would be at least 300 and 125 solar masses (if the low inclination angle is correct). Such high masses cannot explain the observed brightness of R145 if one compares it to other systems with known very high masses such as NGC 3603-A1 and WR20a. So, more data are required to study this potentially very massive and important system.

\citet{heuvel1973} suggested {\bf Cyg X-3} as a final stage of the evolution of a massive star and a neutron star and noted that if the compact star is the black hole, the evolution of the system can follow a similar way. Cyg X-3 is a short period (4.8 hours) binary WR star with a BH candidate, see, e.g., a review by \citet{cherepashchuk2003}. The compact object was assumed as a black hole, and its mass range was estimated as $7M\div40 M_{\odot}$ \citep{schmutz1996}. According to \citet{hanson2000} the mass of the relativistic star is $\lesssim 10 M_{\odot}$ , so it can be either a neutron star or a black hole. \citet{stark2003} found that the mass of the compact object in Cyg X-3 does not exceed $3.6 M_{\odot}$. This fact still allows the possibility that this is a BH+WR system, but there are no final answer about its nature. An additional argument in favour of the hypothesis that there is a BH in the system is that accreting NS with WR stars probably cannot exist, because their rotation should be strongly accelerated during the second episode of the mass transfer, so that they would become either ejectors or propellers \citep{lipunov1982}. The X-ray luminosity of the Cyg X-3 system at 1-60 keV is $\sim 10^{38}$ erg s$^{-1}$, and the bolometric luminosity of the WR star is about $3\cdot 10^{39}$ erg s$^{-1}$. \citet{zdziarski2013} estimated BH mass in the system as $M_{BH}=2.4^{+2.1}_{-1.1} M_{\odot}$, and WR star mass as $M_{WR}=10.3^{+3.9}_{-2.8} M_{\odot}$. Nevertheless, these estimations are only the lower limit of the BH mass \citep{cherepashchuk2013}. \citet{heuvel2017} also have argued that the compact stars in Wolf-Rayet X-ray binaries can only be black holes, and not neutron stars.

{\bf IC 10 X-1} consists of a WR star and potentially one of the most massive stellar BH candidates \citep{prestwich2007}. The mass of the WR star is $32.7\pm 2.6 M_{\odot}$, and the mass of the black hole is $23.1\pm 2.1 M_{\odot}$ \citep{silverman2008}. The orbital period of the system is 34.4 hours. IC 10 X-1 is a bright, variable X-ray source in the metal-poor galaxy IC 10 with a star formation burst. The X-ray luminosity of the system is $\sim 10^{38}$ erg s$^{-1}$ \citep{brandt1997,bauer2004}. The
most probable optical counterpart of the X-ray source is the bright WR star [MAC92] 17A \citep{crowther2003}. The nature of a synchrotron superbubble in IC 10 was studied by \citet{lozinskaya2007} and \citet{lozinskaya2008}. They showed that the most plausible mechanism for the formation of this bubble is a hypernova explosion. IC 10 X-1 is located inside this superbubble, and it is possible that formations of these objects are related.

{\bf NGC 300 X-1} became the third known WR binary with a degenerate companion \citep{caprano2007}. The orbital period of the system is $32.8\pm 0.4$ hours. According to XMM-Newton data, the mean observed
luminosity of the system at 0.2–10 keV is approximately $2\cdot 10^{38}$ erg s$^{-1}$, reaching $10^{39}$ erg s$^{-1}$ (taking into account absorption along the line of sight). According to \citet{crowther2010}, the spectroscopic mass of the optical WN component is $26^{+7}_{-5} M_{\odot}$, implying a mass for the BH of $20\pm 4 M_{\odot}$ for the most likely inclination of the orbit $\approx 60^{\circ}\div 75^{\circ}$.

Recently IC 10 X-1 was studied by \\ \citet{laycock2015a,laycock2015b}, and \\ \citet{steiner2016}; \\ \citet{binder2015} studied NGC 300 X-1.  \\ \citet{laycock2015b} and \\ \citet{binder2015} found that the optical radial velocity curve and X-ray eclipses are not phase connected in IC 10 X-1 and NGC 300 X-1, therefore the dynamical mass estimations of the bodies in these systems remain unreliable. According to \citet{laycock2015a} and \citet{laycock2015b} the compact object in IC 10 X-1 can be even a neutron star. But \citet{steiner2016} studied the spin of the compact object, and using Chandra and NuSTAR data, they argued against a neutron-star model and concluded that IC 10 X-1 contains a black hole of unknown mass.

{\bf SS 433} is the only known supercritical accretor in the Milky Way. SS 433 is a close, eclipsing binary with the orbital period about 13 days (see, e.g., the review \citep{fabrika2004}), in which the donor fills its Roche lobe and the material flows onto a relativistic component in thermal time scale, at the accretion rate about $\sim 10^{-4} M_{\odot}$ yr$^{-1}$ \citep{cherepashchuk1981}. Current estimations of the mass of the compact object do not enable to provide for the final conclusions about its nature. For instance, recently \citet{cherepashchuk2013b} studied X-ray eclipses of SS 433 in hard spectral range and obtained an estimation of the mass of the relativistic companion as $m_x=2.6\div 5.3 M_{\odot}$, the mass of the optical companion was assumed to be $m_v=18 M_{\odot}$.

{\bf M33 X-7} was discovered in the early 1980s \citep{long1981}. The periodic variability of the X-ray source was discovered by \citet{peres1989}. Later, an O6 III optical counterpart of M33 X-7 was found, its minimum mass is $20 M_{\odot}$ and the orbital period of the system is 3.45 d \citep{pietsch2004,pietsch2006}. \citet{orosz2007} estimated masses of relativistic and optical components as $15.65 M_{\odot}$ and $70 M_{\odot}$ correspondingly. \citet{abubekerov2009} found the mass of the compact object to be $15.55\pm 3.20 M_{\odot}$ ( the optical star mass was assumed to be $70 M_{\odot}$), placing it among the most massive stellar BH candidates.

The ``Scenario Machine'' was used by \citet{abubekerov2009}, \citet{bogomazov2014} to study the evolution of M33 X-7, Cyg X-3, IC 10 X-1, NGC 300 X-1, and SS 433. It was shown that the components of M33 X-7 can merge during the second mass exchange stage in the form of the common envelope, and the binary will become a Thorne-Zhytkov object. The other systems are expected to end as double black holes. During the spiral in of the black hole they can be a source of gravitational radiation \citep{nazin1995}. For GW151226 the values of the BH masses $14.2^{+8.3}_{-3.7}M_{\odot}$ and $7.5\pm 2.3M_{\odot}$ \citep{abbott2016d} are very close to values of masses $13.2M_{\odot}$ and $8.41M_{\odot}$ in Figure 2 by \citet{bogomazov2014} for the final double black hole resulting from the Cyg X-3 system. SS 433 in course of its evolution can become a system that consists of a relativistic remnant and a WR star, and finally is able to form a binary relativistic star. Thus, Cyg X-3, IC 10 X-1, and NGC 300 X-1 in the end of their evolution can become binary black holes that can merge within a Hubble time and form gravitational wave sources. During both supernova explosions these systems can become sources of long gamma ray bursts (GRBs) \citep{bogomazov2007,bogomazov2008a,lipunova2009}. The reason is the very fast rotation of nuclei of components in close binaries. The fast rotation prevents the collapse of such nuclei directly into a BH completely, so a part of the matter of such nuclei forms a very compact accretion disc that emits gamma radiation. In general, there are a lot of potential candidates to progenitors of double degenerate binaries and possible sources of gravitational radiation among existing X-ray binaries, see e. g. \citet{laycock2017}.

\section{Scenario Machine}

\label{scenario}

The ``Scenario Machine'' was designed for the population synthesis of the evolution of close binaries.  It can be applied to compute individual evolutionary tracks of close binaries and to investigate properties of groups of close binaries of various types. A very detailed description of this program can be found in a book by \citet{lipunov1996}, and the most recent version by \citet{lipunov2009}, see also a review of the population synthesis methods for modelling the evolution of close binaries by \citet{popov2007}. The first version of this program was created and used by \citet{kornilov1983}. Here we briefly mention only the initial distributions of stellar parameters and parameters of evolution that are free in this study.

We use for the present work the following distribution of the initial semi-major axis $a$:

\begin{equation}
\begin{cases}
\frac{dN}{d\left(\log a\right)} = 0.2,  \\
max\left(10 R_{\odot}, RL\left[M_1\right]\right)\le a \le 10^6 R_{\odot}.
\end{cases}
\label{dista}
\end{equation}

\noindent here $RL\left[M_1\right]$ is the size of the Roche lobe of the primary (initially more massive) star.

The initial component masses are parametrized by a Salpeter mass function:

\begin{equation}
\begin{cases}
f\left(M_1\right)=M_1^{-2.35},  \\
M_{min}\le M_1\le M_{max}.
\end{cases}
\label{distm}
\end{equation}

\noindent here $M_1$ is the initial mass of the primary star, $M_{min}$ and $M_{max}$ are the minimal and the maximum masses of the primary components.

In this work we use four free parameters of stellar evolution: the rate of wind mass loss from stars, the fraction of the mass of the presupernova star that falls under the event horizon during the formation of a BH, the efficiency of the common envelope stage, and the natal kick of a BH.

The stellar mass loss rate $\dot M$ is very important for two reasons: it significantly affects the semi-major axis of the binary and it directly affects the mass of the star itself. We considered the evolutionary scenario A by \citet{lipunov2009}, the mass loss by main-sequence stars is described here using the formula, see e. g. \citep{massevich1979}:

\begin{equation}
\dot M=\frac{\alpha L}{cV_{\infty}}
\label{awind}
\end{equation}

\noindent where $L$ is the star's luminosity, $V_{\infty}$ is the stellar wind velocity at infinity, $c$ is the speed of light, and $\alpha$ is a free parameter. In scenario A, the decrease in the star's mass $\Delta M$ does not exceed 10\% of its hydrogen envelope during one evolutionary stage. We parametrized the mass loss by WR stars as

\begin{equation}
\Delta M_{WR}=\alpha_{WR}M_{WR},
\label{awindwr}
\end{equation}

\noindent here $M_{WR}$ is the initial stellar mass in the WR stage.

The mass of a BH $M_{BH}$ formed by an exploding presupernova of mass $M_{preSN}$, was calculated as

\begin{equation}
M_{BH}=k_{bh} M_{preSN},
\label{kbh}
\end{equation}

\noindent where the coefficient $k_{bh}$ is the fraction of the presupernova mass that forms the BH. 

During the common envelope stage binary stars very efficiently lose their angular momentum with the lost envelope matter, and the components approach one another along a spiral trajectory. The efficiency of mass loss in the common envelope stage is described by the parameter $\alpha_{CE}=\Delta E_b/\Delta E_{orb}$, where $\Delta E_b=E_{grav}-E_{thermal}$ is the binding energy of the ejected envelope and $\Delta E_{orb}$ is the reduction in the orbital separation during the approach:

\begin{equation}
\alpha_{CE}\left(\frac{GM_a M_c}{2a_f}-\frac{GM_a M_d}{2a_i}\right)=\frac{GM_d (M_d - M_c)}{R_d},
\label{ace}
\end{equation}

\noindent here $M_c$ is the core mass of the mass losing star with initial mass $M_d$ and radius $R_d$ (this is a function of the initial semi-major axis $a_i$ and initial component mass ratio $M_a/M_d$, where $M_a$ is the mass of the accretor), $a_f$ is the final semi-major axis in the end of the common envelope stage.

The common envelope forms if the Roche lobe overflow occurs\footnote{According to Kippenhahn and Weigert classification and Webbink diagram. See the description of the code by \citet{lipunov2009} for more details, subsection 4.4 ``RL Filling''.} in a type C system (where the star that fills its Roche lobe has a strongly evolved core), the whole mass ratio range, even for $q\sim 1$. For type B systems, we use the condition $q\le q_{cr} = 0.3$ for the formation of the common envelope, otherwise the systems evolves without the common envelope. \citet{heuvel2017}\footnote{See also \citep{pavlovskii2017}.} studied very similar conditions as for type B systems for the common envelope formation for the study of WR$+$O stars and came to a conclusion that BH$+$BH merger rate is $\sim 10^{-5}$ per year in a galaxy like Milky Way, confirming the results obtained using the ``Scenario Machine'' \\ by \citet{bogomazov2008a}, Table 2. 

An additional natal kick velocity can be acquired by BHs in this form:

\begin{equation}
v_{BH}=v_a\frac{M_{preSN}-M_{BH}}{M_{BH}},
\label{kickbh}
\end{equation}

\noindent where $v_{BH}$ is the natal kick velocity of a BH, $v_a$ is a parameter that is distributed as

\begin{equation}
f(v_a) \sim \frac{v_{a}^2}{v_0^3} e^{-\frac{v_{a}^2}{v_0^2}},
\label{vanis}
\end{equation}

\noindent and $v_0$ is a free parameter.

\section{Population synthesis}

\label{results}

This paper is dedicated to the most massive merging BHs with masses $M_{BH1} \geq 25 M_{\odot}$ and $M_{BH2} \geq 25 M_{\odot}$. Masses of the most massive stars can reach up to several hundred of solar masses, e. g. \citet{popescu2014}, but masses of components even in the most massive known binaries don't exceed $\approx 150$ solar masses (see Section \ref{black-wolf}). In population synthesis studies it's usually accepted that stars with initial mass $\ge 20-25 M_{\odot}$ produce a BH in the end of their evolution\footnote{However, see also \citet{sukhbold2016}, who showed that black holes may form from even less massive stars, and neutron stars can still form from quite massive stars.}. Pair instability pulsations can destroy the star during its supernova explosion without a remnant \citep{fryer2001b,heger2002,belczynski2016c}. Therefore we use three models: (i) the initial star's mass (or the maximum mass of the star in the course of its evolution) that produces a BH is $\ge 25 M_{\odot}$ and it does not have an upper limit, (ii) WR stars which have masses before the explosion in the range $65 M_{\odot}\le M_{WR}\le 135 M_{\odot}$ explode without a remnant, other WR stars with initial (or maximum) masses $\ge 25 M_{\odot}$ produce a BH, (iii) the initial mass of the stars that produce BHs under our consideration is $\le 140 M_{\odot}$ (and $\ge 25 M_{\odot}$). Additional mass loss due to pair instability pulsations (see, e. g. a study by \citet{yoshida2016}) without the disruption of the collapsing star is assumed as negligible in comparison with the mass loss due to the stellar wind.

\begin{figure*}
\includegraphics[width=1.0\textwidth]{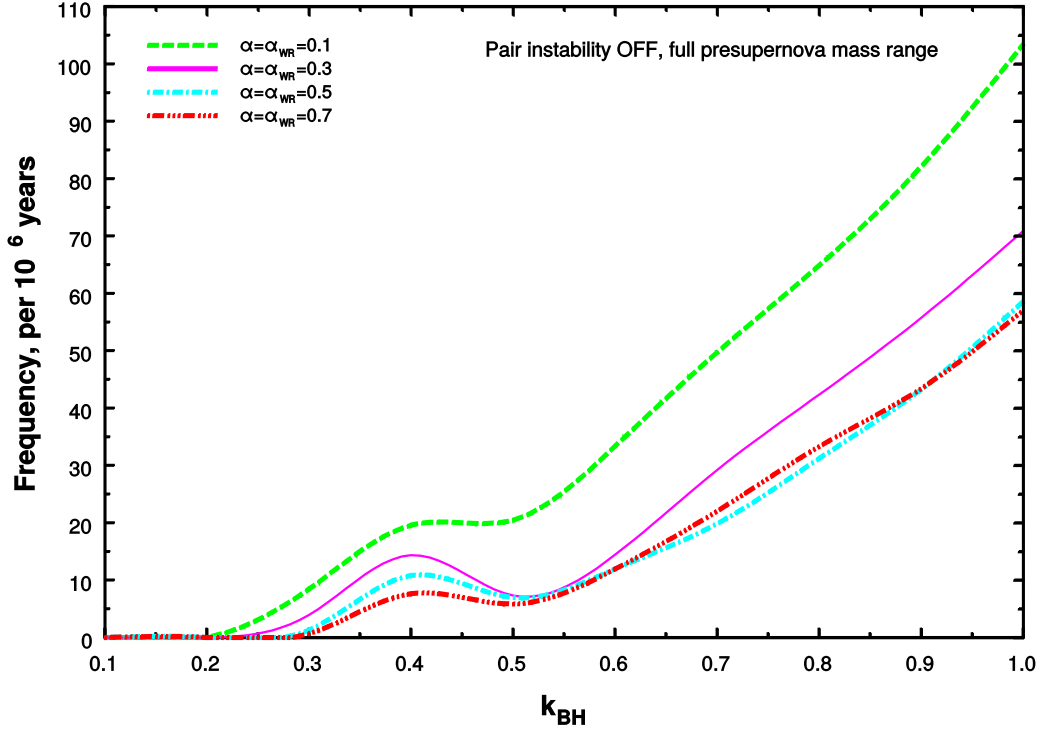}
\vspace{15pt} \caption{Merger frequencies (per galaxy like the Milky Way) of two black holes with masses $M_{BH1} \ge 25 M_{\odot}$ and $M_{BH2} \ge 25 M_{\odot}$ as a function of the part of the presupenova mass $k_{BH}$ that falls under the event horizon, here $M_{BH1}$ and $M_{BH2}$ are masses of merging black holes. The curves correspond to different values of the mass loss by the stellar wind (parameters $\alpha$ and $\alpha_{WR}$). The assumed common envelope efficiency is $\alpha_{CE}=0.5$, and the black hole kick is $v_0=0$. Initial and presupernova massess are not limited, model (i).}\label{f1}
\end{figure*}

\begin{figure*}
\includegraphics[width=1.0\textwidth]{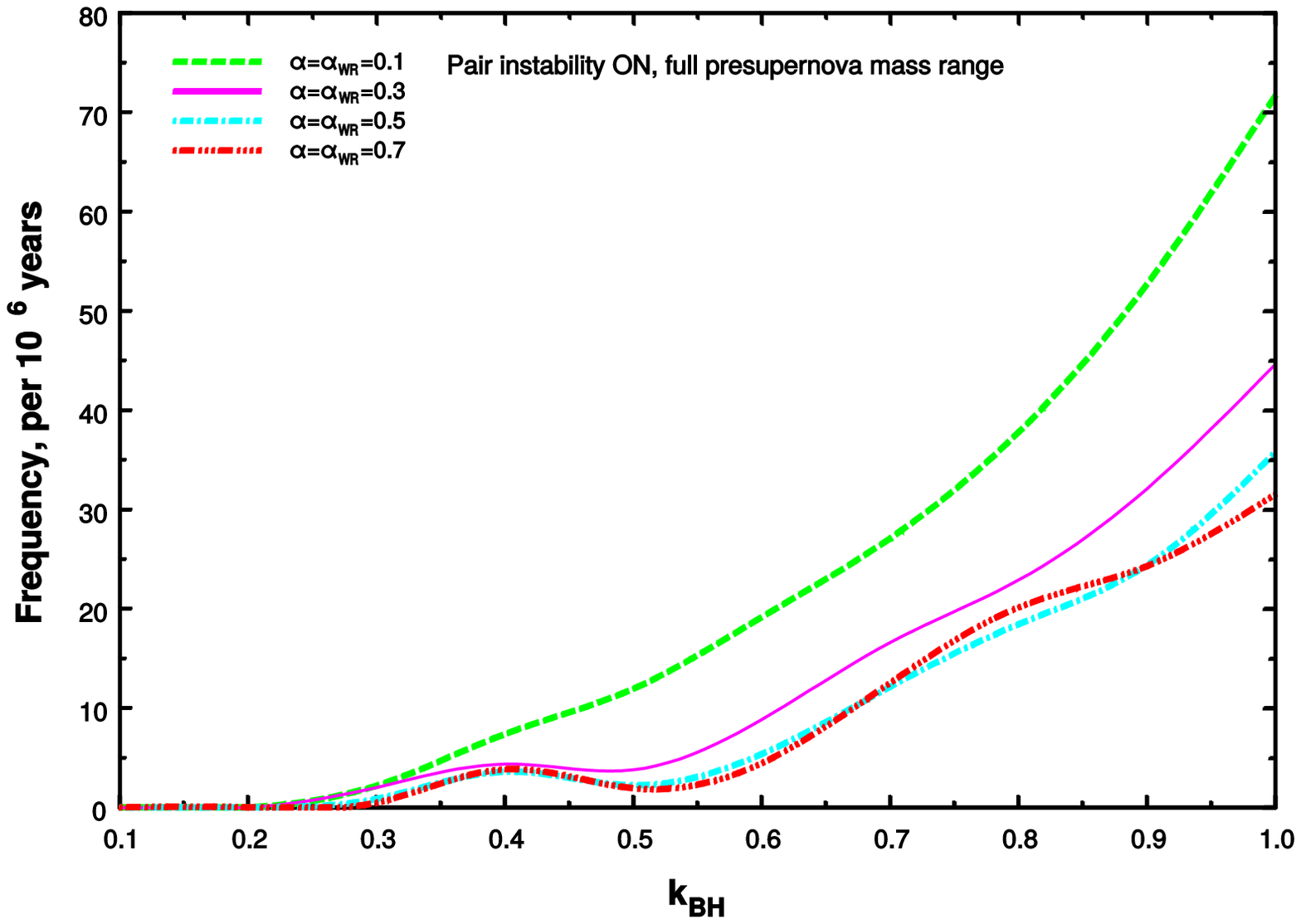}
\vspace{15pt} \caption{The same as Figure \ref{f1}. Wolf-Rayet stars with presupernova masses in the range $65 M_{\odot} \le M_{WR} \le 135 M _{\odot}$ are assumed to explode without leaving a remnant due to pair instability in their cores, model (ii).}\label{f2}
\end{figure*}

\begin{figure*}
\includegraphics[width=1.0\textwidth]{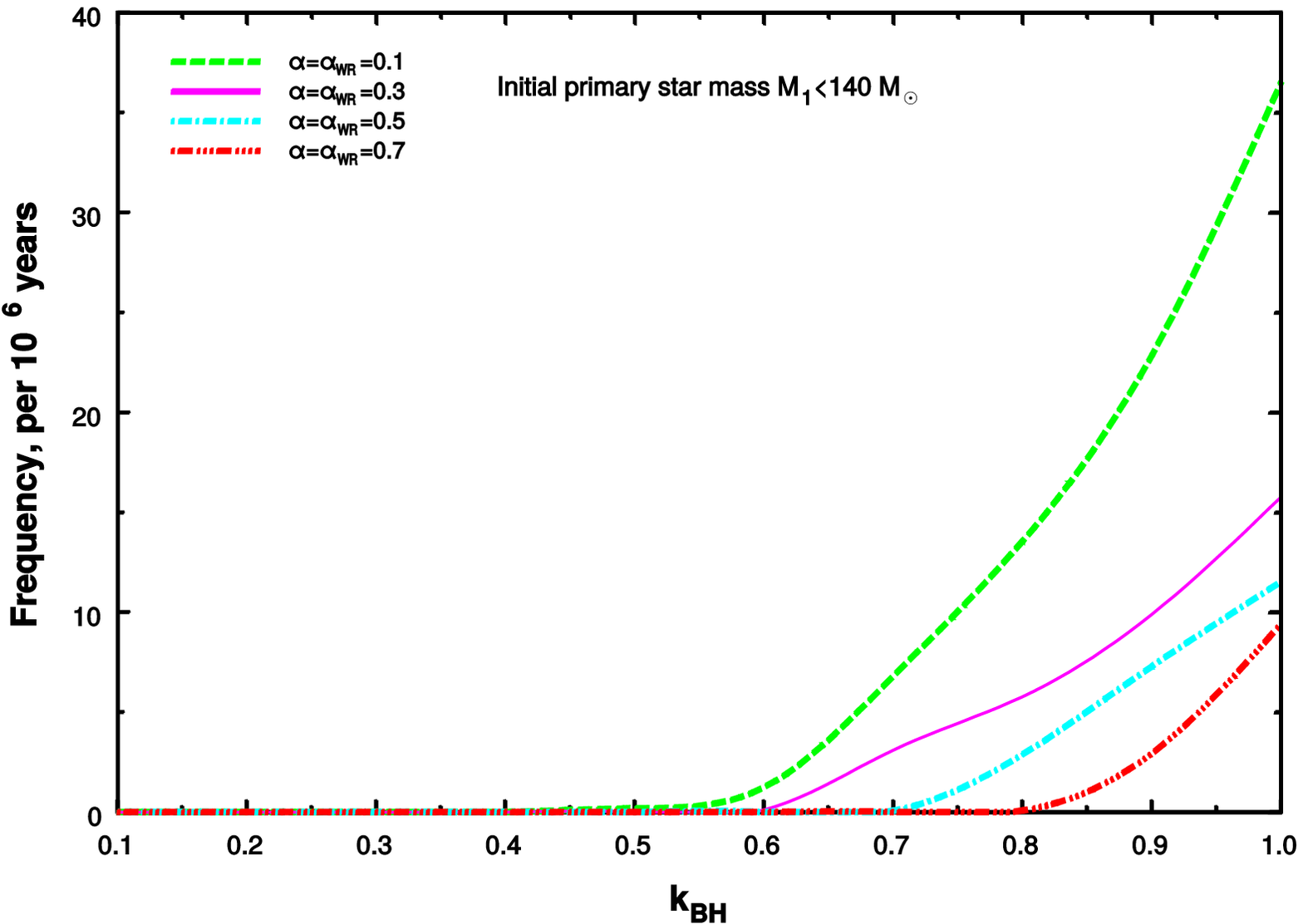}
\vspace{15pt} \caption{The same as Figure \ref{f1}. The primary star initial mass is assumed to be less than $140 M _{\odot}$, model (iii).}\label{f3}
\end{figure*}

\begin{figure*}
\includegraphics[width=1.0\textwidth]{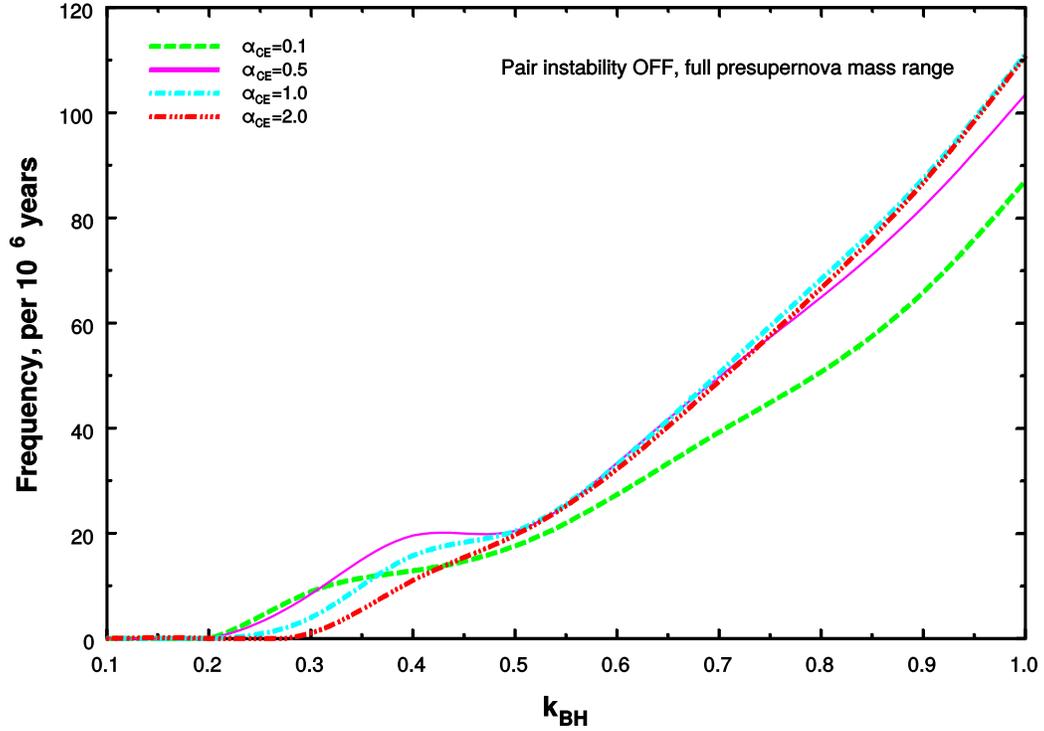}
\vspace{15pt} \caption{Merger frequencies (per galaxy like Milky Way) of two black holes with masses $M_{BH1} \ge 25 M_{\odot}$ and $M_{BH2} \ge 25 M_{\odot}$ depending on the part of the presupenova mass $k_{BH}$ that falls under the event horizon, here $M_{BH1}$ and $M_{BH2}$ are masses of merging black holes. The curves correspond to the different values of the common envelope efficiency $\alpha_{CE}$. Stellar wind mass loss parameters $\alpha=0.1$ and $\alpha_{WR}=0.1$, and the black hole kick is $v_0=0$. Initial and presupernova massess are not limited, model (i).}\label{f4}
\end{figure*}

\begin{figure*}
\includegraphics[width=1.0\textwidth]{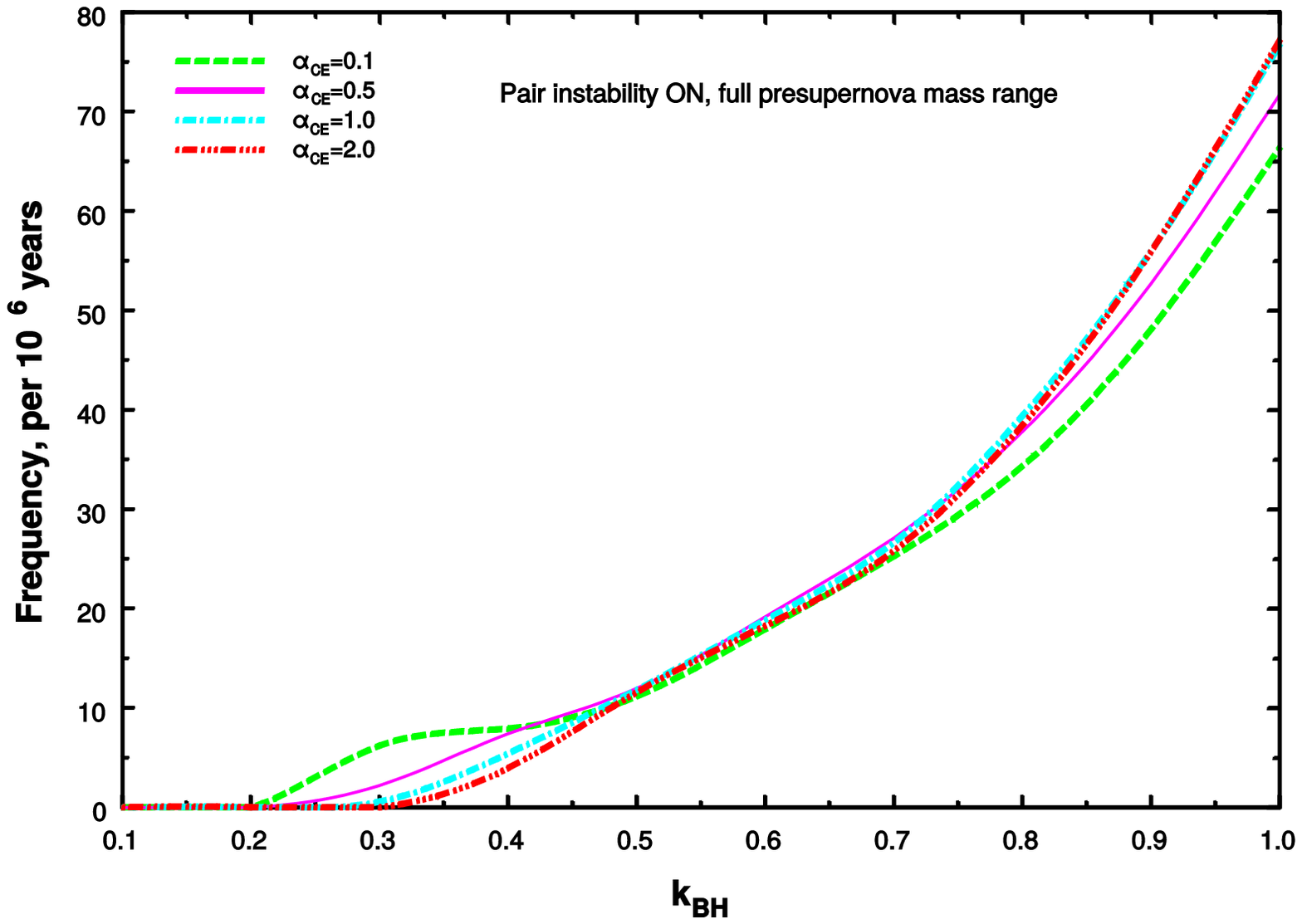}
\vspace{15pt} \caption{The same as Figure \ref{f1}. Wolf-Rayet stars with presupernova masses in the range $65 M_{\odot} \le M_{WR} \le 135 M _{\odot}$ are assumed to explode without leaving a remnant due to pair instability in their cores, model (ii).}\label{f5}
\end{figure*}

\begin{figure*}
\includegraphics[width=1.0\textwidth]{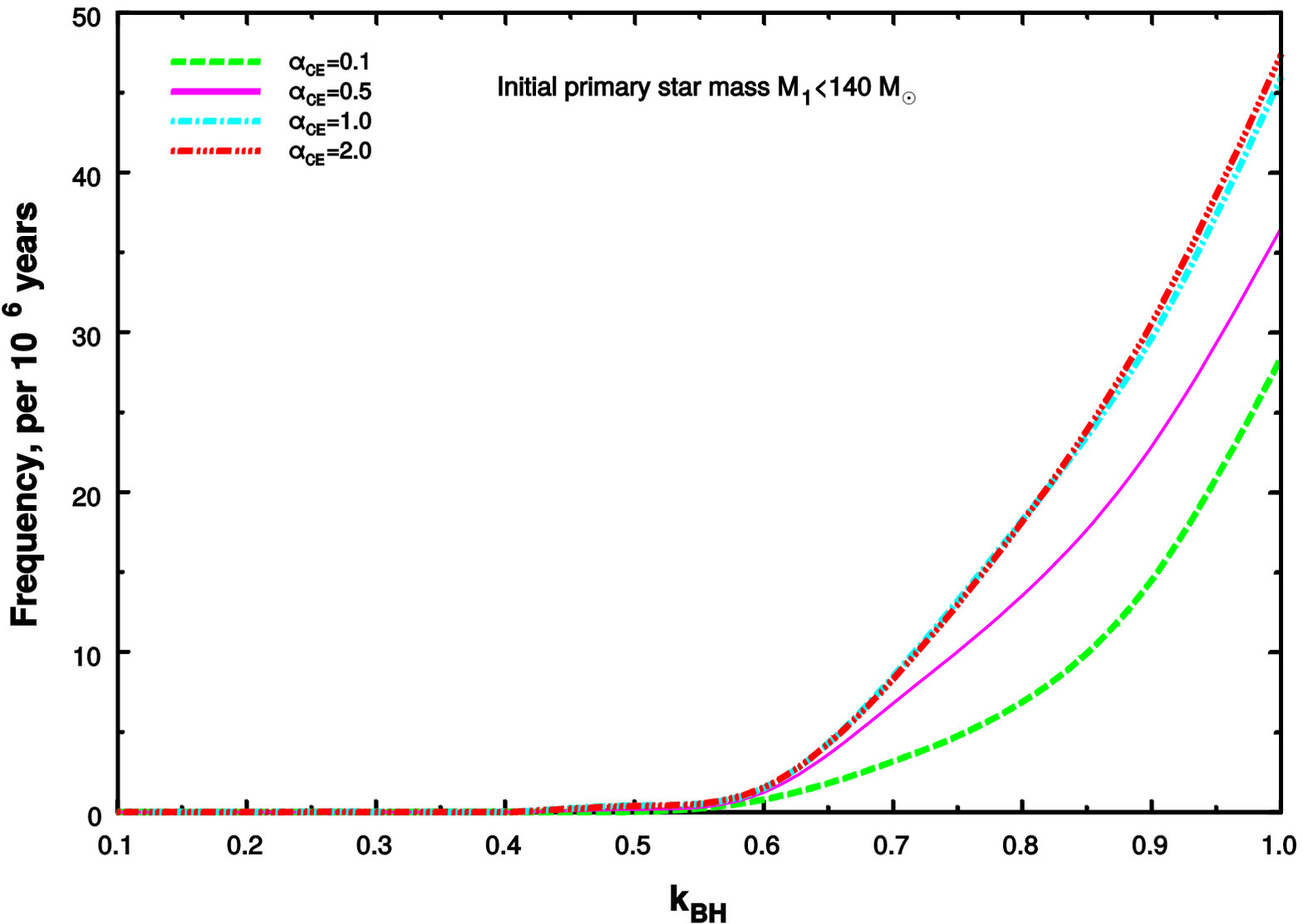}
\vspace{15pt} \caption{The same as Figure \ref{f4}. The primary star initial mass is assumed to be less than $140 M _{\odot}$, model (iii).}\label{f6}
\end{figure*}

\begin{figure*}
\includegraphics[width=1.0\textwidth]{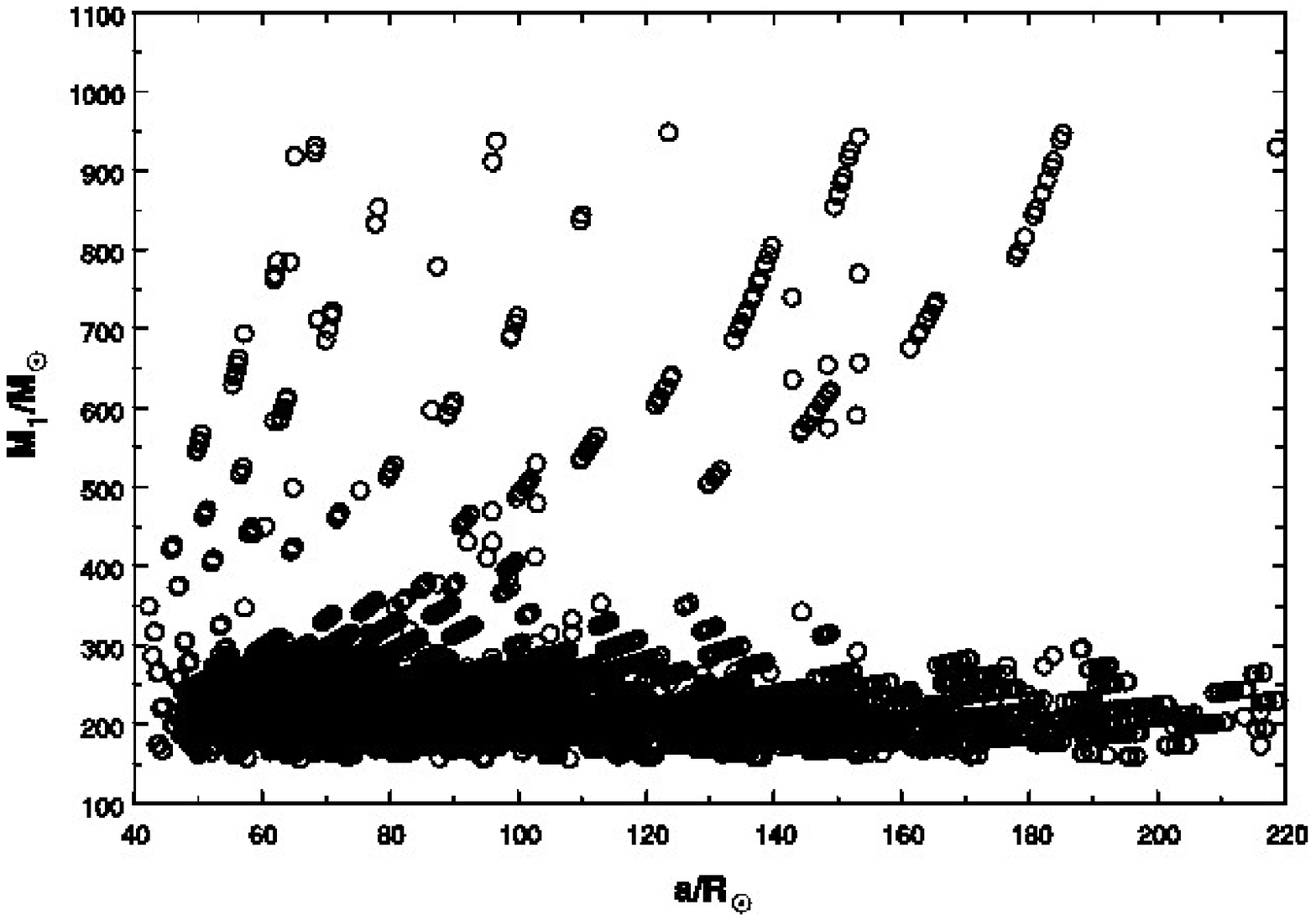}
\vspace{15pt} \caption{Initial masses of primary components versus initial semi-major axises in the zero age main sequence. The evolution of these systems leads to the formation of merging black holes with masses $M_{BH1} \ge 25 M_{\odot}$ and $M_{BH2} \ge 25 M_{\odot}$. $k_{bh}=0.35$, $\alpha_{CE}=0.5$, $\alpha=\alpha_{WR}=0.1$, $v_0=0$. Model (i). }\label{f7}
\end{figure*}

\begin{figure*}
\includegraphics[width=1.0\textwidth]{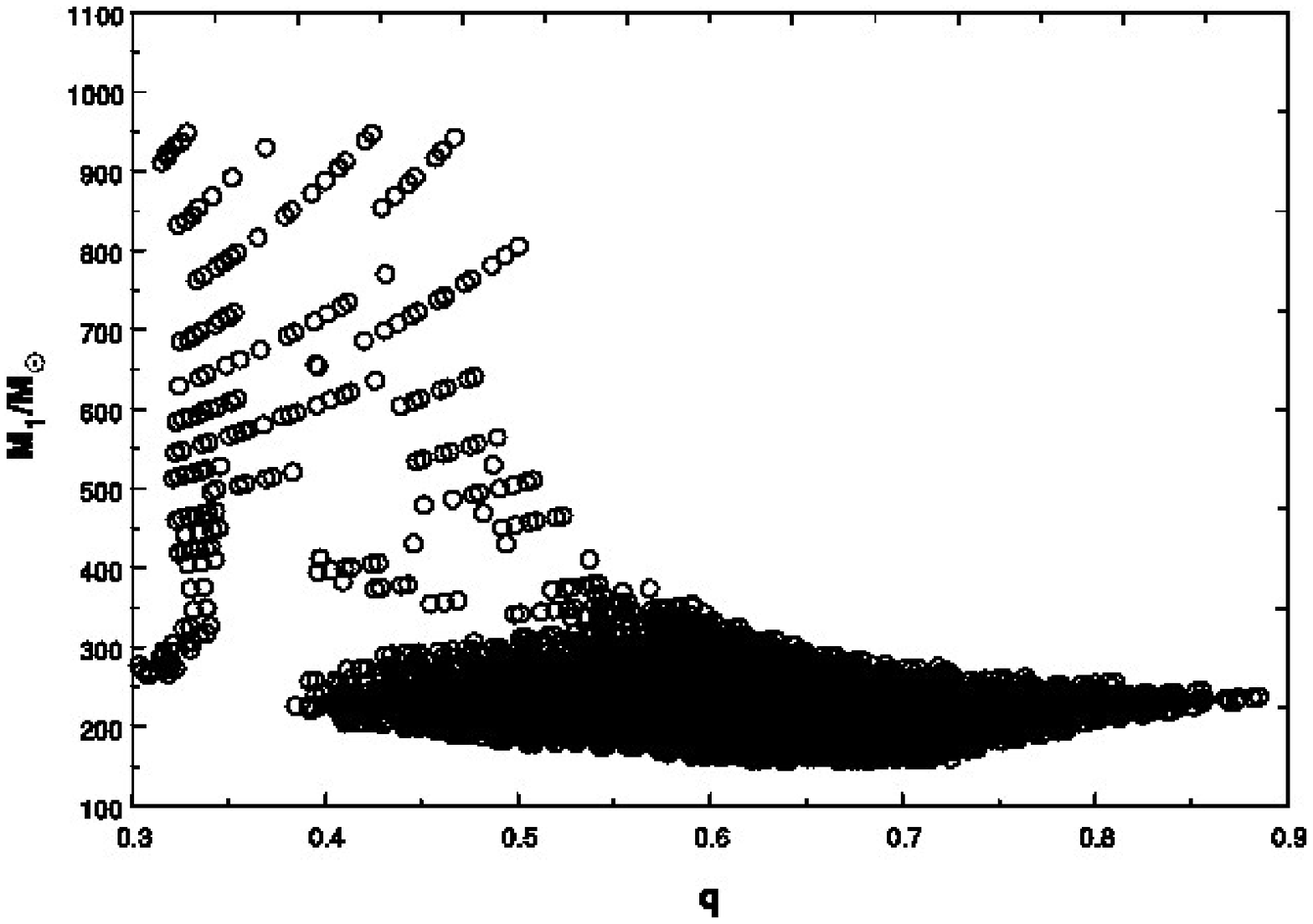}
\vspace{15pt} \caption{Initial masses of primary components versus initial mass ratio $q=M_2/M_1\le 1$ on the zero age main sequence. The evolution of these systems leads to the formation of merging black holes with masses $M_{BH1} \ge 25 M_{\odot}$ and $M_{BH2} \ge 25 M_{\odot}$, for the assumed parameters:$k_{bh}=0.35$, $\alpha_{CE}=0.5$, $\alpha=\alpha_{WR}=0.1$, $v_0=0$.  Model (i). }\label{f8}
\end{figure*}

\begin{figure*}
\includegraphics[width=1.0\textwidth]{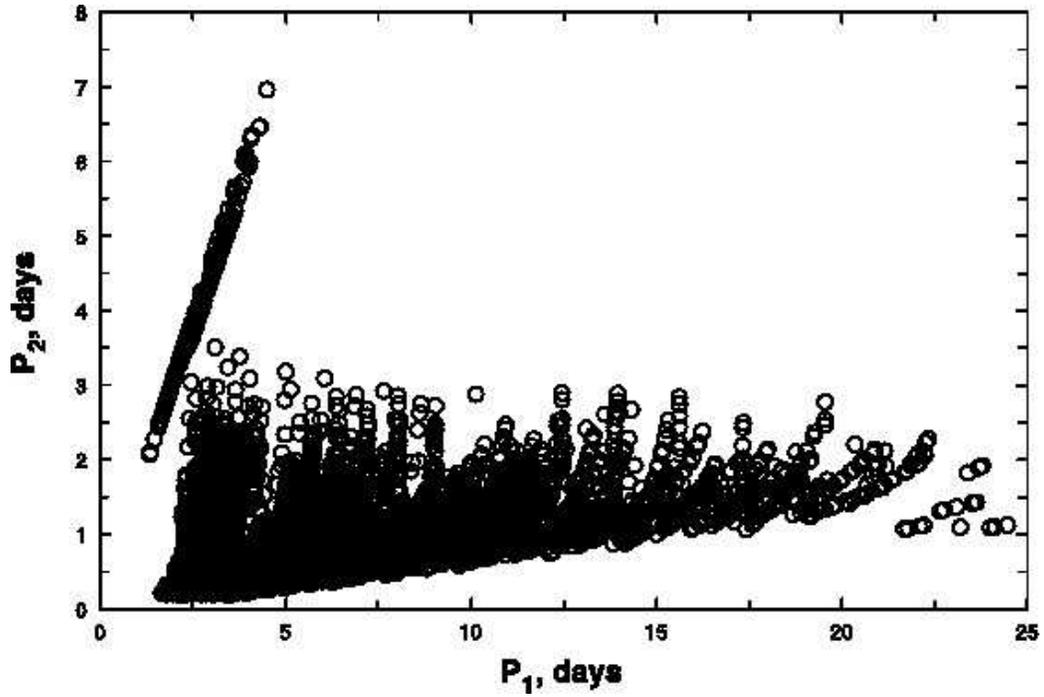}
\vspace{15pt} \caption{ $\text{P}_2$ versus $\text{P}_1$ diagram. $\text{P}_1$ is the binary's orbital period before the first supernova explosion, $\text{P}_2$ is the binary's orbital period before the second supernova explosion. Assumed parameters are: $k_{bh}=0.35$, $\alpha_{CE}=0.5$, $\alpha=\alpha_{WR}=0.1$, $v_0=0$. The evolution of these systems leads to the formation of merging black holes with masses $M_{BH1} \ge 25 M_{\odot}$ and $M_{BH2} \ge 25 M_{\odot}$.  Model (i). }\label{f9}
\end{figure*}

\begin{figure*}
\includegraphics[width=1.0\textwidth]{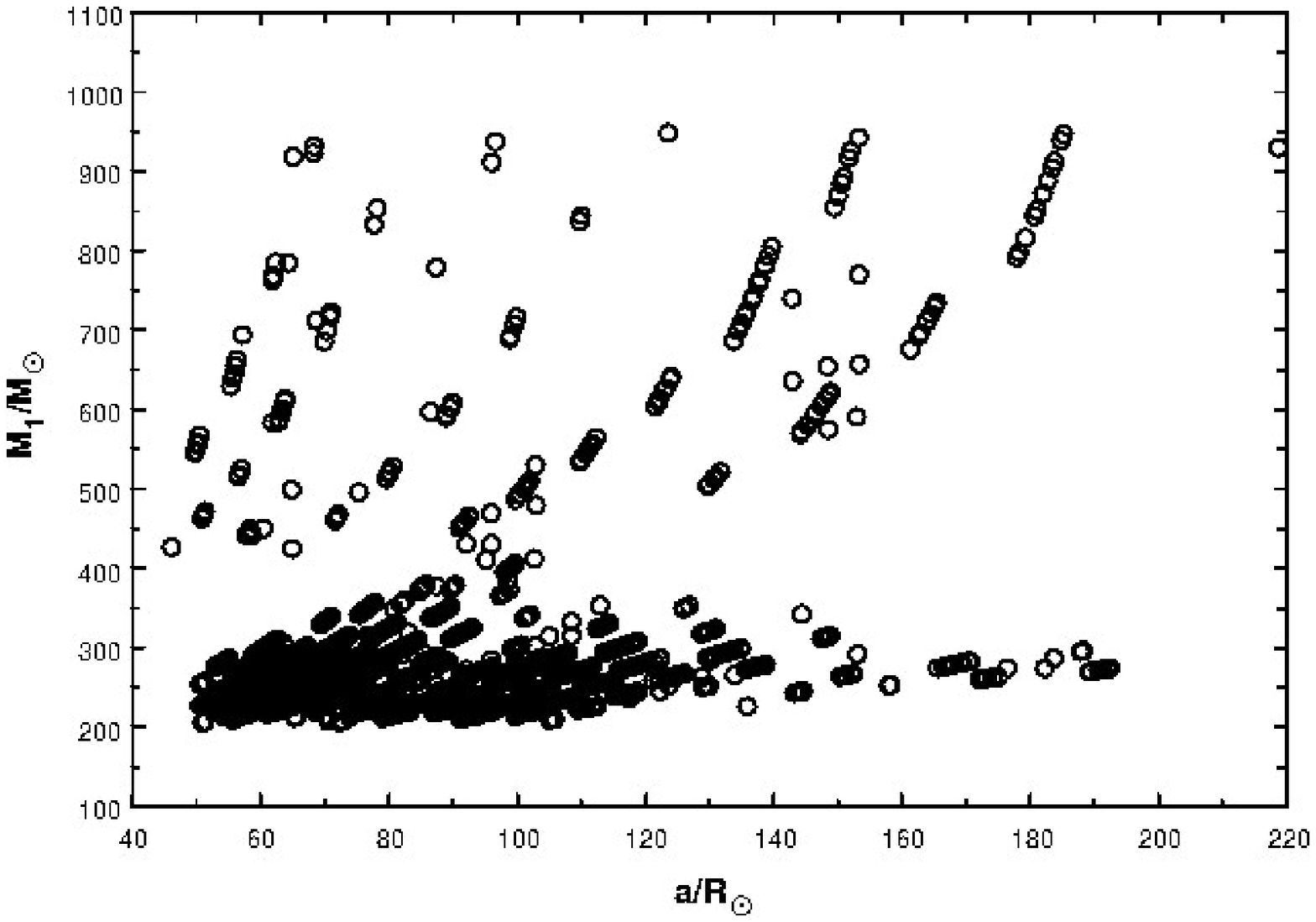}
\vspace{15pt} \caption{The same as Figure \ref{f7} for the model (ii).}\label{f7a}
\end{figure*}

\begin{figure*}
\includegraphics[width=1.0\textwidth]{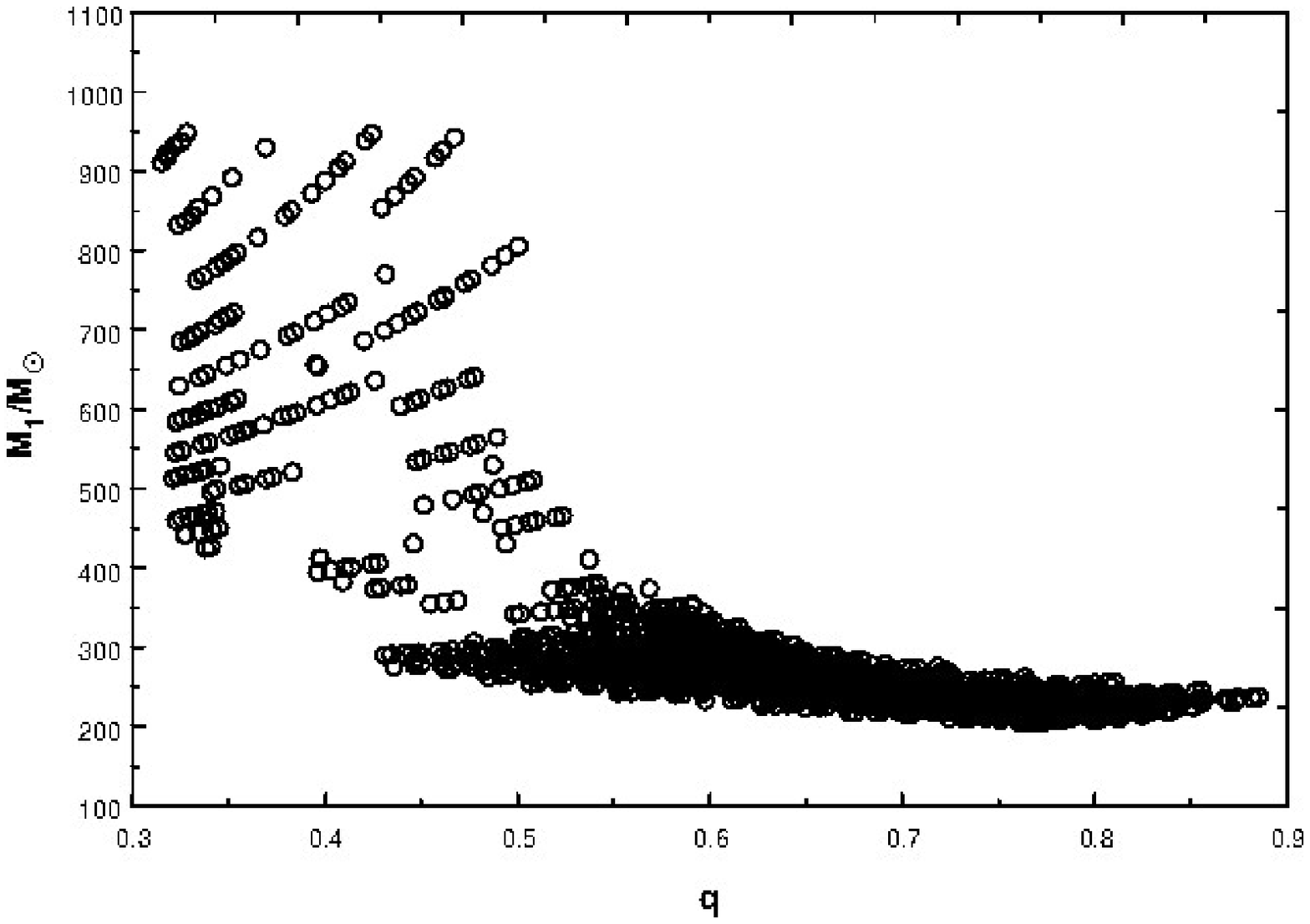}
\vspace{15pt} \caption{The same as Figure \ref{f8} for the model (ii).}\label{f8a}
\end{figure*}

\begin{figure*}
\includegraphics[width=1.0\textwidth]{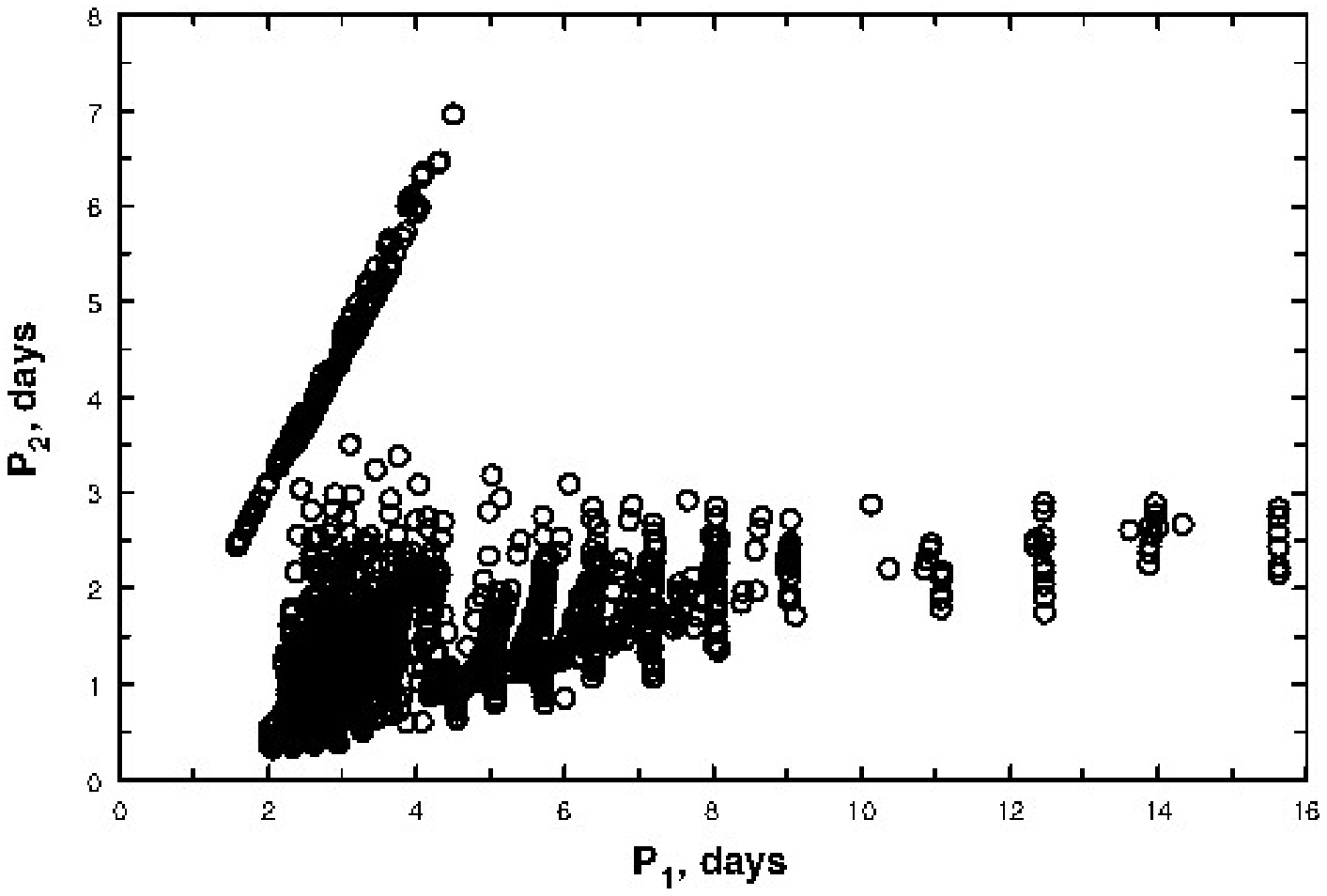}
\vspace{15pt} \caption{The same as Figure \ref{f9} for the model (iii).}\label{f9a}
\end{figure*}

\begin{figure*}
\includegraphics[width=1.0\textwidth]{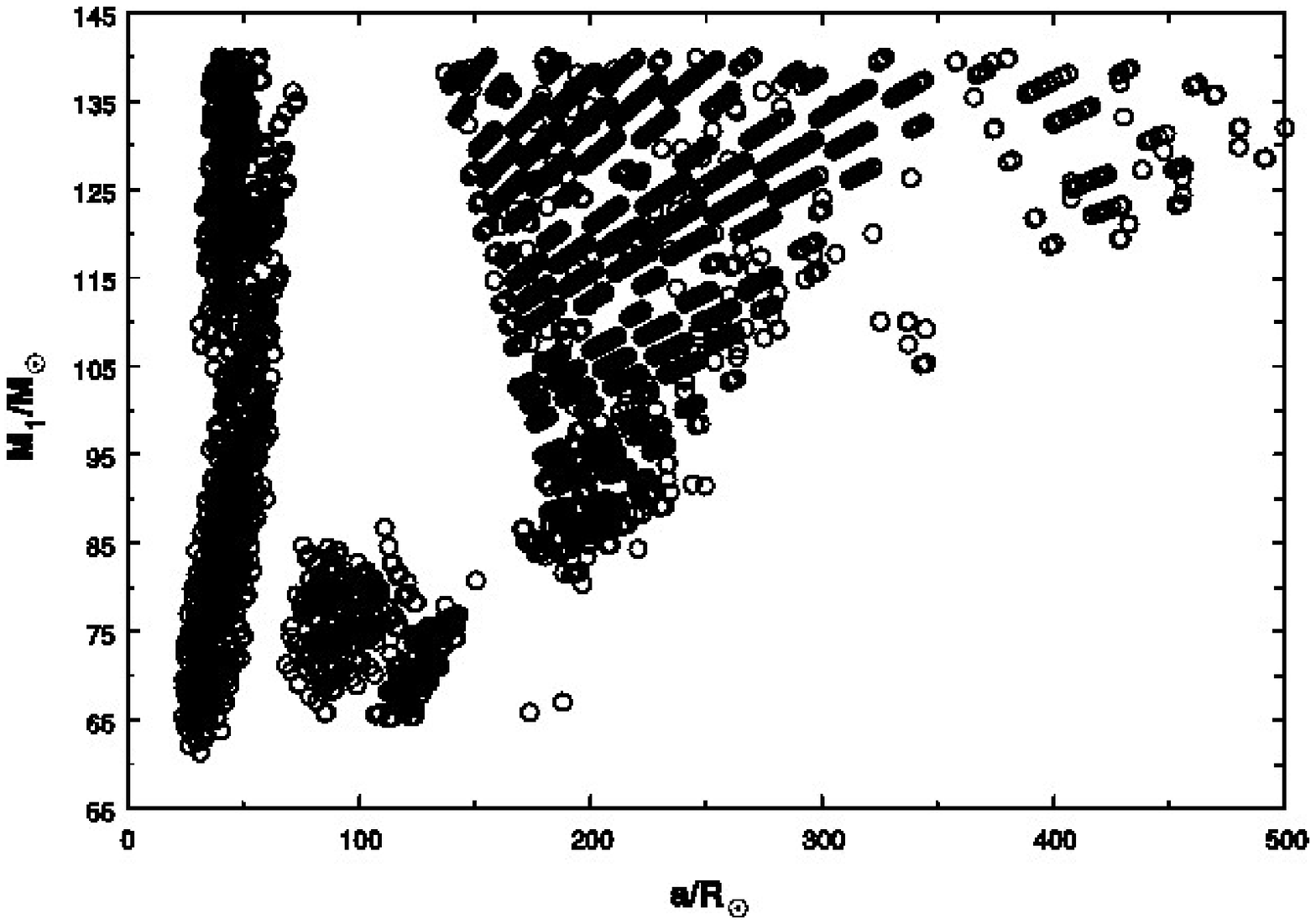}
\vspace{15pt} \caption{The same as Figure \ref{f7} for $k_{bh}=0.8$ and model (iii).}\label{f10}
\end{figure*}

\begin{figure*}
\includegraphics[width=1.0\textwidth]{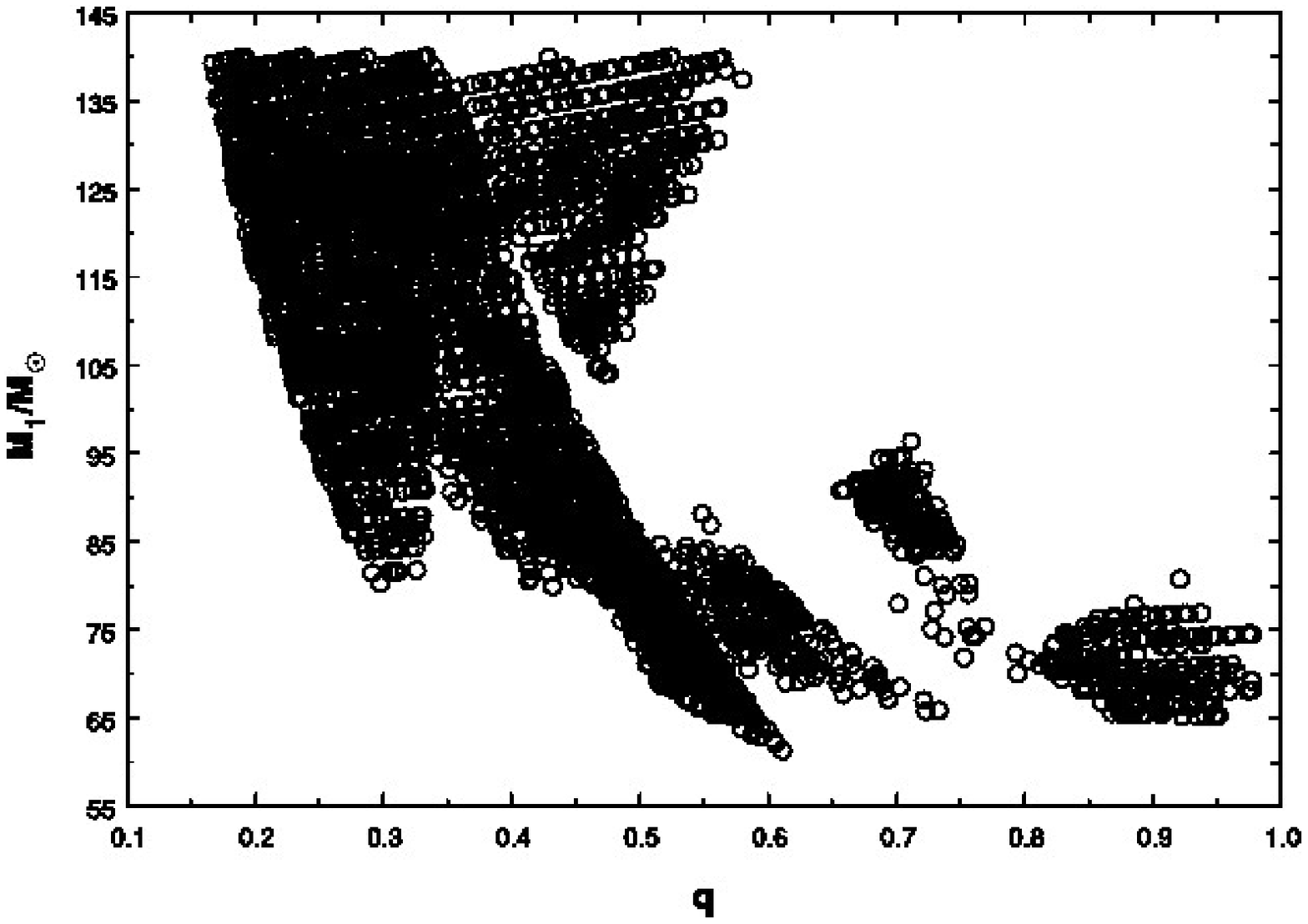}
\vspace{15pt} \caption{The same as Figure \ref{f8} for $k_{bh}=0.8$ and model (iii).}\label{f11}
\end{figure*}

\begin{figure*}
\includegraphics[width=1.0\textwidth]{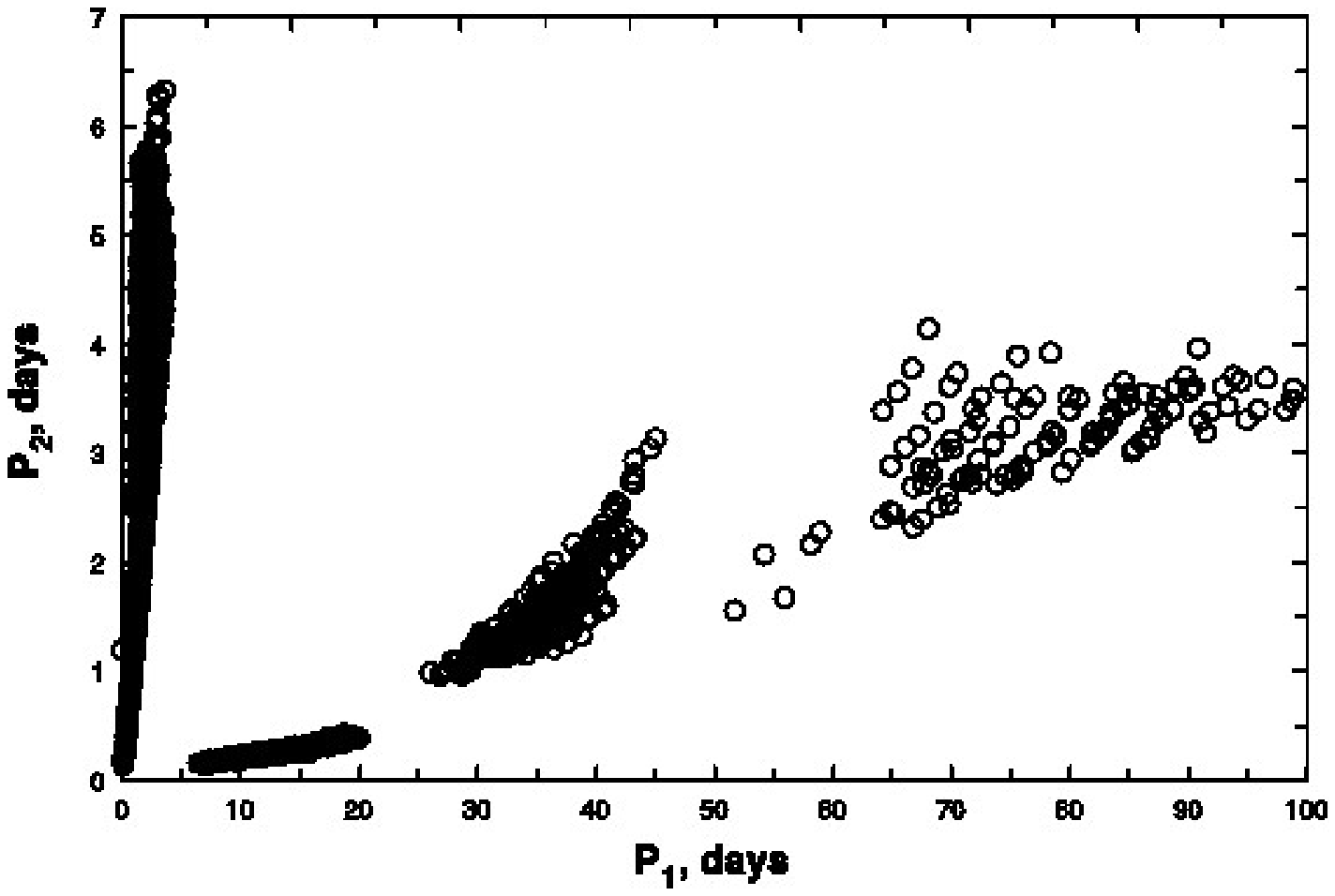}
\vspace{15pt} \caption{The same as Figure \ref{f9} for $k_{bh}=0.8$ and model (iii).}\label{f12}
\end{figure*}

\begin{figure*}
\includegraphics[width=1.0\textwidth]{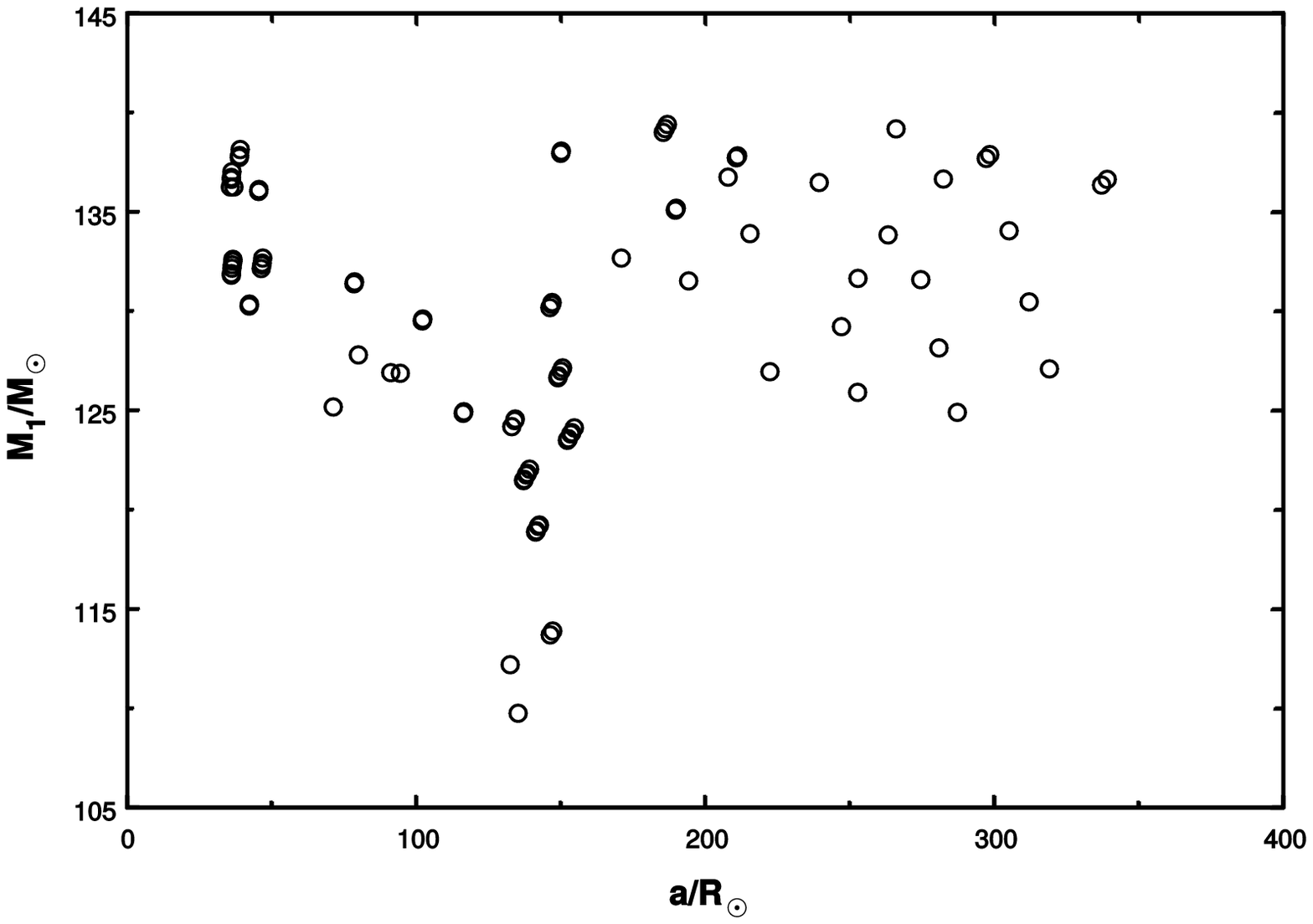}
\vspace{15pt} \caption{The same as Figure \ref{f7} for $k_{bh}=0.55$ and model (iii).}\label{f13}
\end{figure*}

\begin{figure*}
\includegraphics[width=1.0\textwidth]{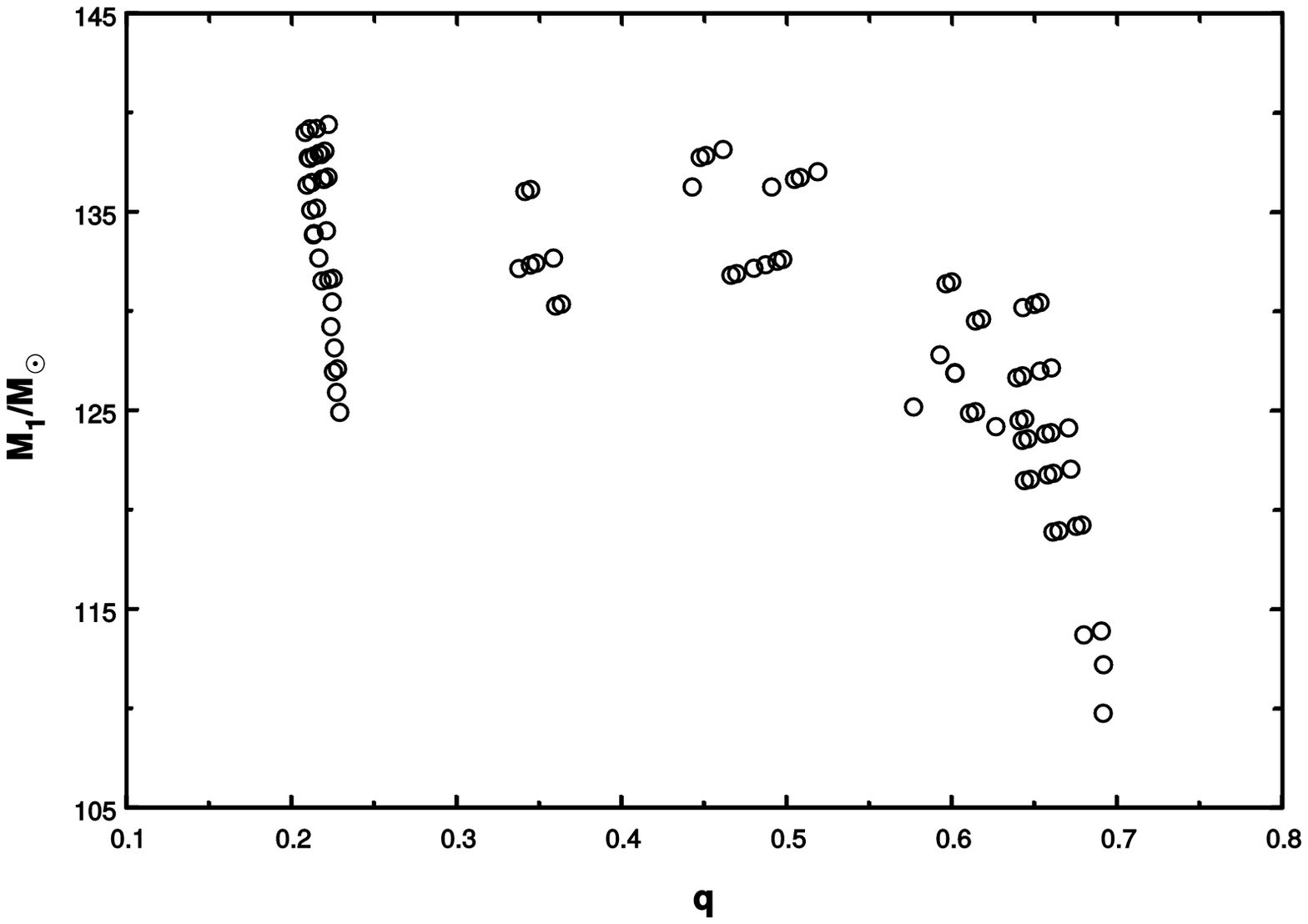}
\vspace{15pt} \caption{The same as Figure \ref{f8} for $k_{bh}=0.55$ and model (iii).}\label{f14}
\end{figure*}

\begin{figure*}
\includegraphics[width=1.0\textwidth]{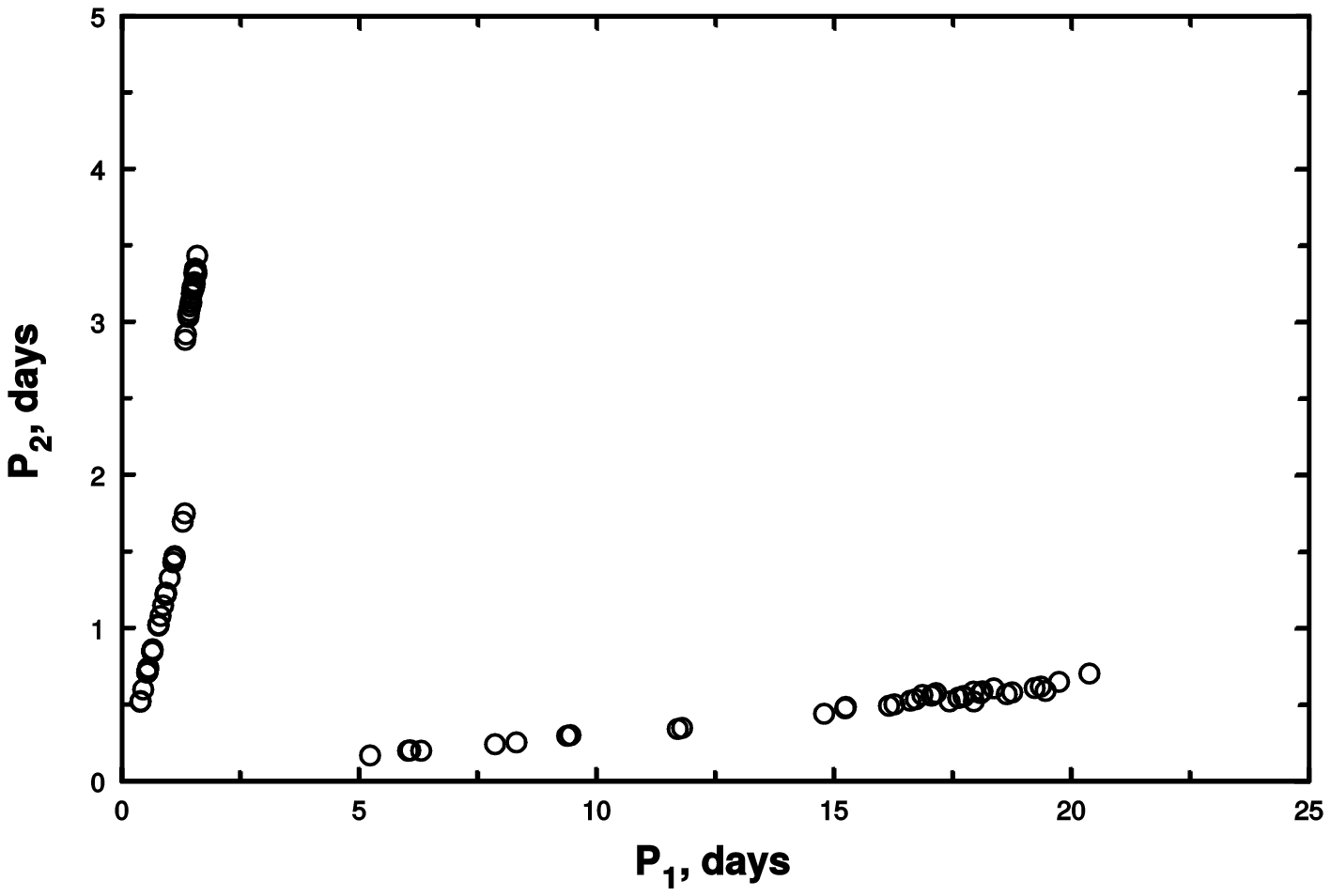}
\vspace{15pt} \caption{The same as Figure \ref{f9} for $k_{bh}=0.55$ and model (iii).}\label{f15}
\end{figure*}

\begin{figure*}
\includegraphics[width=1.0\textwidth]{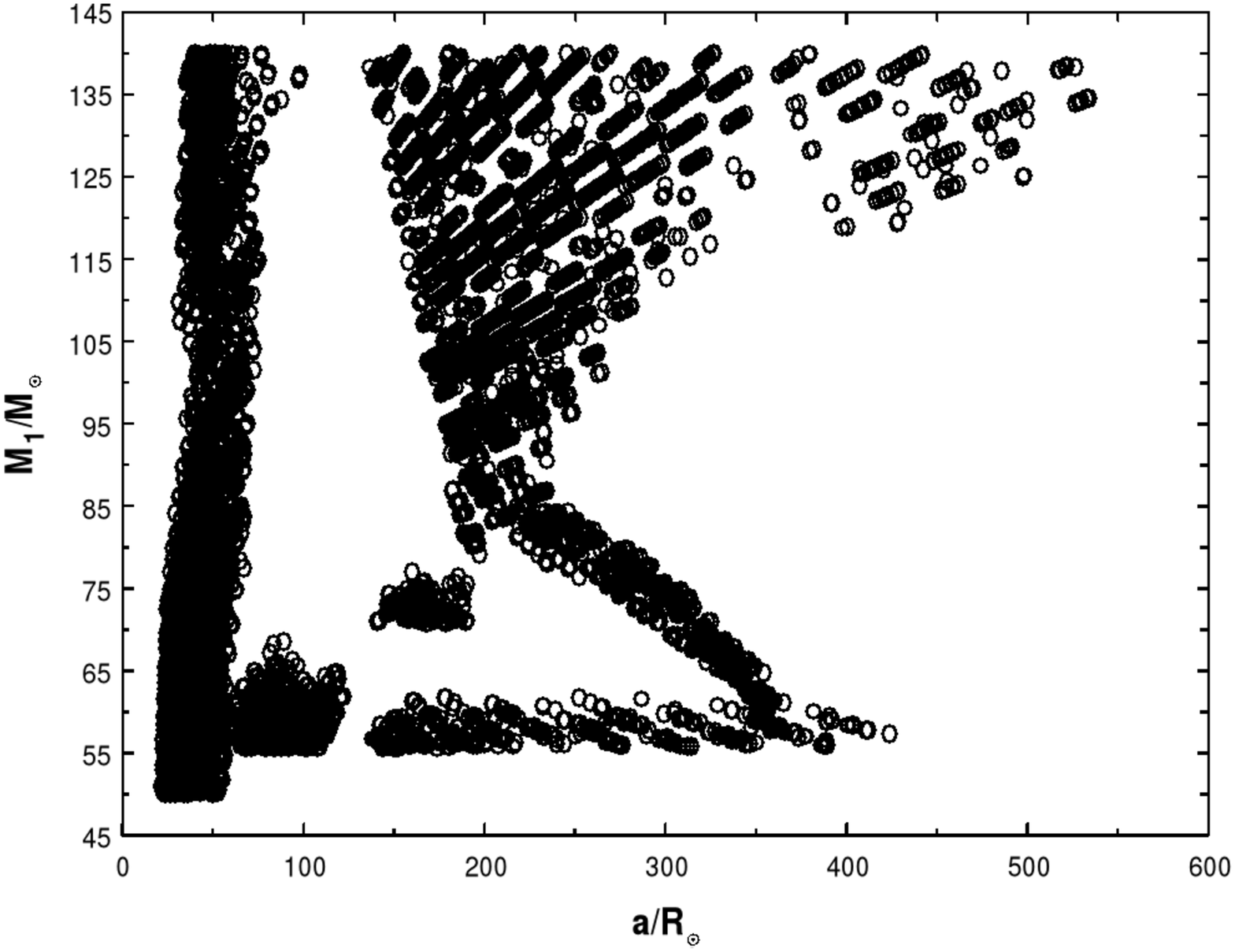}
\vspace{15pt} \caption{The same as Figure \ref{f7} for $k_{bh}=1.0$ and model (iii).}\label{f16}
\end{figure*}

\begin{figure*}
\includegraphics[width=1.0\textwidth]{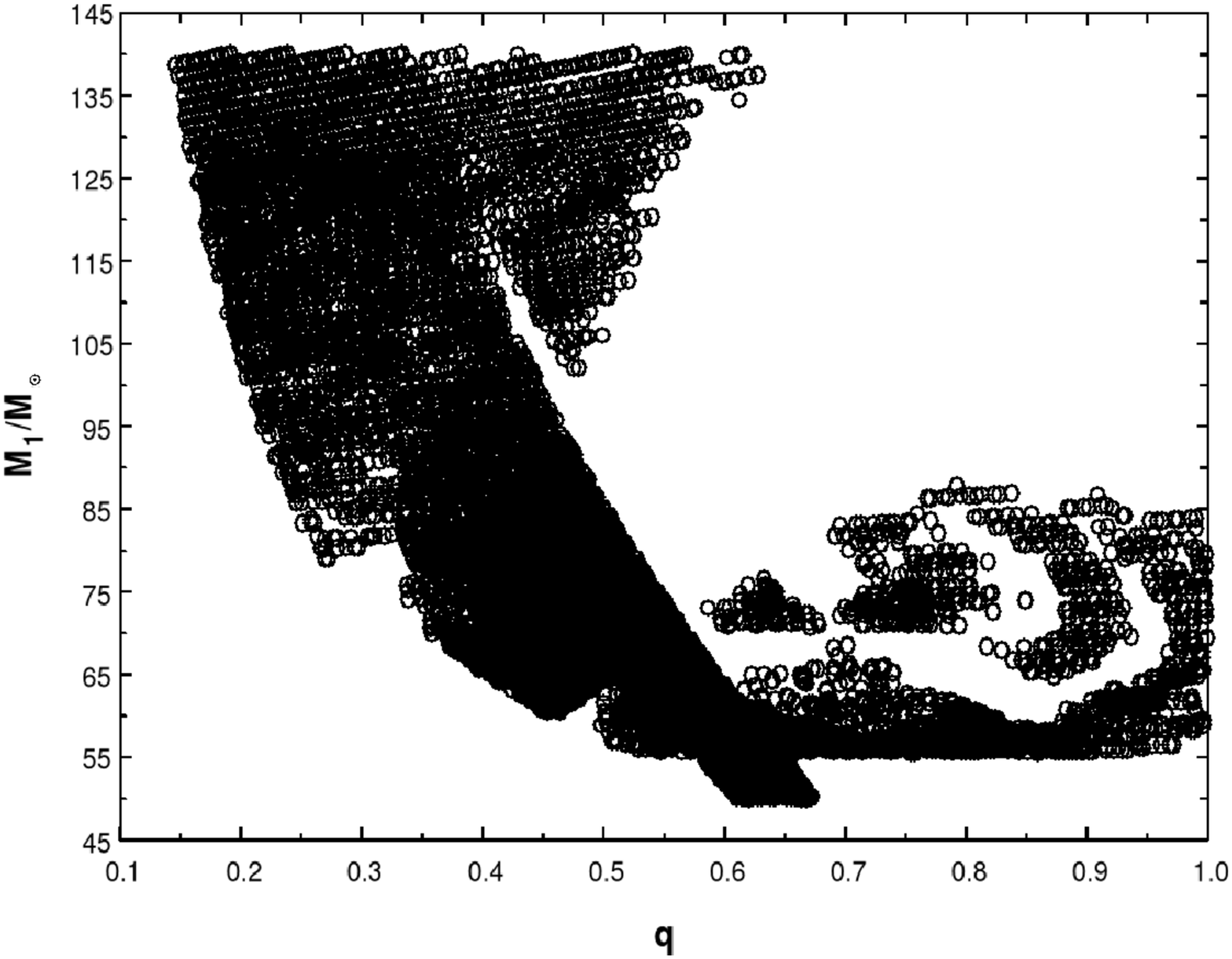}
\vspace{15pt} \caption{The same as Figure \ref{f8} for $k_{bh}=1.0$ and model (iii).}\label{f17}
\end{figure*}

\begin{figure*}
\includegraphics[width=1.0\textwidth]{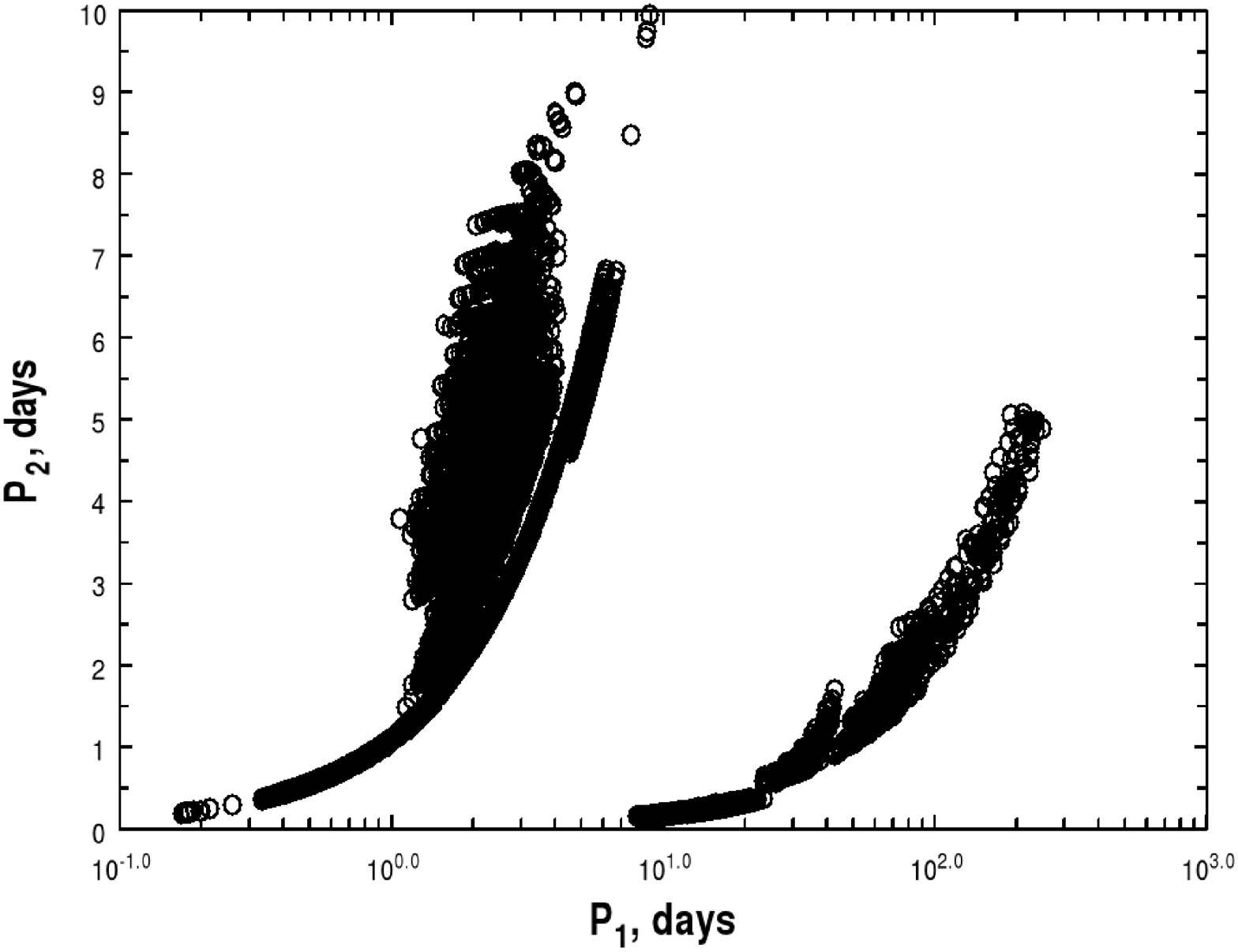}
\vspace{15pt} \caption{The same as Figure \ref{f9} for $k_{bh}=1.0$ and model (iii).}\label{f18}
\end{figure*}

\begin{figure*}
\includegraphics[width=1.0\textwidth]{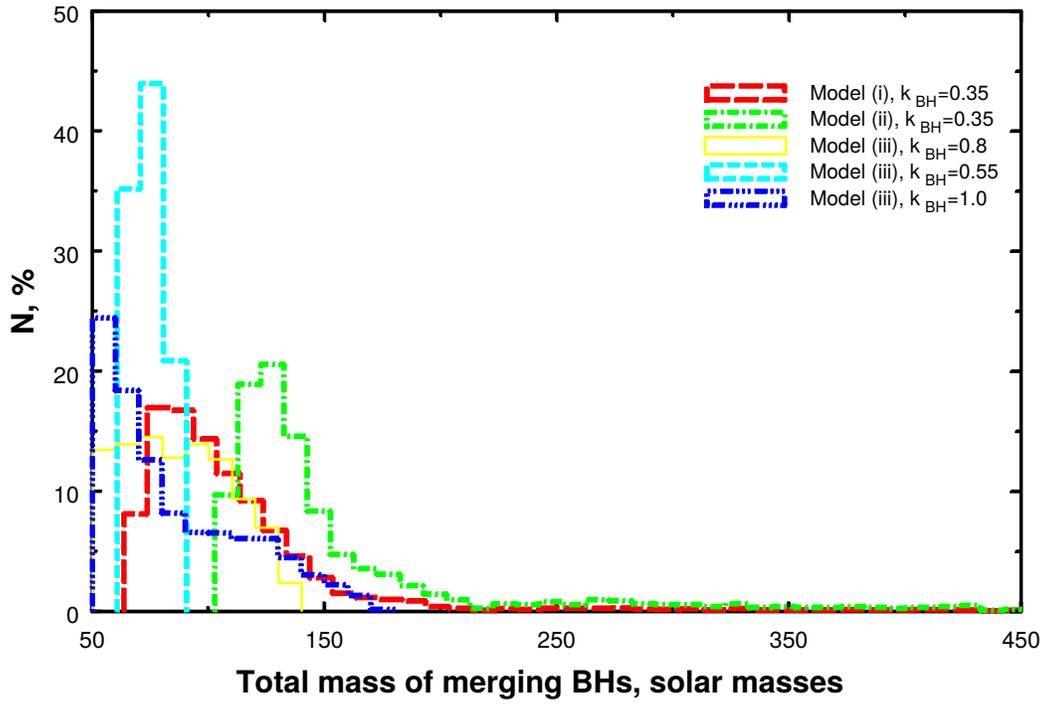}
\vspace{15pt} \caption{ Distribution of masses $M_{BH1}+M_{BH2}$ of merging BHs for different models and different values of $k_{BH}$ ($M_{BH1} \ge 25 M_{\odot}$ and $M_{BH2} \ge 25 M_{\odot}$). }\label{f19}
\end{figure*}

\begin{figure*}
\includegraphics[width=1.0\textwidth]{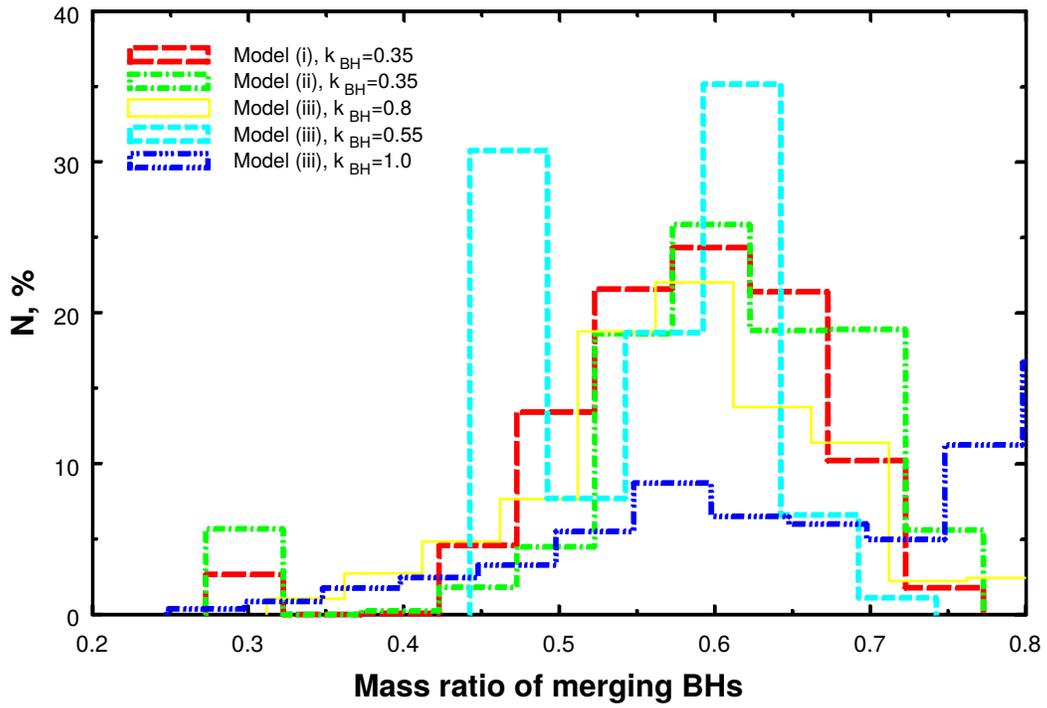}
\vspace{15pt} \caption{Mass ratio distribution for merging BHs for different models and different values of $k_{BH}$ ($M_{BH1} \ge 25 M_{\odot}$ and $M_{BH2} \ge 25 M_{\odot}$). }\label{f20}
\end{figure*}

The results of our calculations are presented in Figures \ref{f1}-\ref{f20}, and in Tables \ref{track1}-\ref{kerr}. We took $M_{min}=50M_{\odot}$, and $M_{max}=950M_{\odot}$ in Equation (\ref{distm}) following the upper limit of masses of stars by \citet{popescu2014}, the initial mass ratio $q=M_2/M_1\le 1$ of binary components was assumed as equiprobable (i.e. a flat distribution). For each set of initial parameters, a population synthesis was performed for $10^6$ binaries. Free parameters in our calculations take the following values: $\alpha=$ 0.1, 0.3, 0.5, 0.7, $\alpha_{WR}=$0.1, 0.3, 0.5, 0.7, $\alpha_{CE}=$0.1, 0.3, 0.5, 0.7, 1.0, 2.0, $k_{bh}=$0.1, 0.2, 0.3, 0.4, 0.5, 0.6, 0.7, 0.8, 0.9, 1.0, $v_0=$0, 50, 100 km s$^{-1}$.

In Tables \ref{track1}-\ref{track4} we collected as examples several evolutionary tracks that lead to a merger of BHs with the appropriate masses found in our computations. The track in the Table \ref{track1} does not contain a CE stage, in RL stages conditions for CE formation are not met. The track in the Table \ref{track2} represents a variant, where the CE stage occurs before the first supernova explosion, and after it CE is not able to form. And Table \ref{track3} represents a track, where the binary goes through a CE stage before the second supernova explosion, whereas before the first explosion a CE does not form. Table \ref{track4} represents a track with masses of the components nearer to each other than in Table \ref{track2}, so both components become WR stars before the first explosion, leaving the possibility of almost equal orbital periods before the first and the second explosions. The evolutionary channel from Table \ref{track4} is rare and realizes only under the assumption of very specific sets of free parameters. These two sets are: (a) $\alpha_{CE}=0.1\div 2.0$, $v_0=0\div 100$ km s$^{-1}$, $\alpha=0.1$, $\alpha_{WR}=0.1$, $k_{BH}=0.9\div 1.0$; (b) $\alpha_{CE}=0.1\div 2.0$, $v_0=0\div 100$ km s$^{-1}$, $\alpha=0.7$, $\alpha_{WR}=0.7$, $k_{BH}=0.8\div 1.0$. The BH+BH merger frequency for such kind of tracks is $1\div 1.5 \cdot 10^{-6}$ per year in a galaxy like the Milky Way in both sets. Results for other sets of parameters are given in Tables \ref{track1}-\ref{track3}.

In Figures \ref{f1}-\ref{f6} we show the merger frequencies of BHs under our consideration depending on free scenario parameters. All frequencies are normalized per a galaxy like the Milky Way. From these figures we can conclude:

\begin{itemize}

\item The weaker is the stellar wind, the higher is the merger frequency of BHs under investigation.

\item The merger rate of massive BHs weakly depends on the CE stage efficiency $\alpha_{CE}$.

\item The merger frequency of BHs with appropriate masses increases when the coefficient $k_{BH}$ increases.

\end{itemize}

If $v_o$ becomes significant ($\approx 50$ km s$^{-1}$ and higher), most BHs of the appropriate masses almost do not merge. This statement is valid only for the BHs under consideration in this paper and potentially can be an artifact of the code.

In Figures \ref{f7}-\ref{f18} we present initial $M_1$--$a$, $M_1$--$q$ distributions of the studied systems, and $P_1$--$P_2$ diagrams. $P_1$ is the orbital period of the binary at the moment preceding the first supernova explosion, $P_2$ is the same just before the second explosion. For these plots we used five sets of parameters, for all these sets we assumed $\alpha_{CE}=0.5$, $\alpha=\alpha_{WR}=0.1$, and $v_0=0$:

\begin{enumerate}[a.]

\item Model (i), $k_{bh}=0.35$.

\item Model (ii), $k_{bh}=0.35$.

\item Model (iii), $k_{bh}=0.8$.

\item Model (iii), $k_{bh}=0.55$.

\item Model (iii), $k_{bh}=1.0$.

\end{enumerate}

As we can see from Figures \ref{f7} and \ref{f8} the initial mass in case (a) is $100\div 900 M_{\odot}$, the semi-major axis is $40-220 R_{\odot}$, initial mass ratio is in the rang $0.3-0.9$. In case (b) these quantities are almost the same, only a part of systems disapper from the figures in comparison with (a) due to disruption of the stars by the pair instability, see Figures \ref{f7a} and \ref{f8a}. In case (c) $M_1\approx 57\div 140M_{\odot}$, $a\approx 20\div 500R_{\odot}$, and $q\approx 0.2\div 1.0$ (Figures \ref{f10} and \ref{f11}). In case (d) $M_1\approx 105\div 140M_{\odot}$, $a\approx 20\div 400R_{\odot}$, and $q\approx 0.2\div 0.7$ (Figures \ref{f13} and \ref{f14}). In case (e) $M_1\approx 50\div 140M_{\odot}$, $a\approx 15\div 550R_{\odot}$, and $q\approx 0.1\div 1.0$ (Figures \ref{f16} and \ref{f17}). Figures \ref{f9}, \ref{f9a}, \ref{f12}, \ref{f15}, and \ref{f18} shows that the effective Kerr parameters of two merging BHs can vary in a wide range and could be met in different combinations practically independently of scenario parameters that were accepted as free parameters in the present study, see Section \ref{rotation}.

Figures \ref{f19} and \ref{f20} depict distributions of merging BHs under consideration on total masses and on mass ratios for future comparisons with experimental data.

\section{Effective Kerr parameter}

\label{rotation}

In the Table \ref{kerr} we study the effective dimensionless Kerr parameter before SN explosions:

\begin{equation}
a_K=\frac{I\Omega}{GM^2_c/c},
\label{kerreq}
\end{equation}

\noindent where $I$ is the moment of inertia of the core, $\Omega$ is the angular velocity and $M_c$ is the core mass. The radius of the core by the end of helium burning is taken from the mass and temperature using the virial theorem:

\begin{equation}
R_c=\frac{G\mu m_p M_c}{6kT},
\label{hcore}
\end{equation}

\noindent where $R_c$ is the core radius, $T$ is the temperature
of carbon burning, which is about $6\cdot 10^8$ K, $G$ is the gravitational constant, $\mu=15$ is the average number of nucleons for a particle, $m_p$ is the proton mass, and $k$
is the Boltzmann constant. The moment of inertia can be written as $I = k_I M_c R^2_c$ , where a
dimensionless parameter $k_I$ takes values 0.4 for a uniform spherical body and 0.1 for polytrope spheres with the polytropic index $n=2.5$. We assume here that in the closest binaries the CO core of a star is fully synchronized with the orbital rotation and then evolves without significant angular momentum losses untill the collapse (final pre-supernova rotates faster than the synchronous rate). In general this assumption can be valid only for the most close binaries, at the same time this approach allows to estimate maximum possible spin of a BH formed in a binary. From the Table \ref{kerr} we see that the core of the collapsing WR star can possess enough rotational momentum to produce a GRB, and probably the spin of the BH formed in the explosion with a GRB has a value close to one. If the evolution of a track goes though the CE stage the collapsing core of the WR star in the first supernova explosion probably does not possess enough rotational momentum for a gamma ray burst, and therefore the spin of a BH can be much less than one. The maximum value of the semi-major axis of a binary BHs with masses equal to those of LIGO GW150914 (with merging time less than the age of the Universe) is $a\approx 50\div 60 R_{\odot}$, see Equation 20 by \citet{lipunov2009}. This quantity corresponds to the orbital period approximately equal to 4-6 days leaving the possibility of spin $\ll 1$ for both merging BHs.

\section{Discussions and conclusions}

\label{final}

An evolutionary track of GW150914 was calculated by \citet{woosley2016} and by \citet{belczynski2016b}. According to \citet{woosley2016} the delayed merger of two black holes in the end of the evolution of a close binary is the most probable model for making the gravitational radiation burst. \citet{belczynski2016b} also found that the existence of GW150914 does not require enhanced double black hole formation in dense stellar clusters or in other exotic evolutionary channels. The observed dimensionless spin magnitude for the primary BH is $a_1= 0.31^{+0.48\pm 0.04}_{-0.28\pm 0.01}$ and for the secondary $a_2= 0.46^{+0.48\pm 0.07}_{-0.42\pm 0.01}$, and upper limits for these quantities are $a_1=0.69\pm 0.05$ and $a_2=0.88\pm 0.1$, see Table 1 by \citet{ligo2016}. \citet{kushnir2016} estimated the merger time as $>10^8$ years using assumptions of low spin. All tracks in Tables \ref{track1}--\ref{track4} meet this requirement. According to our calculations the Kerr parameter $a$ can take any value lower than one, depending on their formation history and the mass concenration to the center (parameter $k_I$) of a collsapsing stellar core. It is essential to note that BHs that have grown primarily through accretion probably are not maximally rotating \citep{gammie2004}. \citet{belczynski2016b} also suggest that the spin of merging BHs is mostly the natal spin. The SN explosion of the closest binaries during their collapse can be accompanied with a long gamma ray burst \citep{woosley1993}, see also \citet{lipunova2009}. The possibility of a GRB during a definite collapse strongly depends on the rotation evolution of the collapsing core. The final BH spin as well depends on the angular momentum loss by the collapsing matter. This evolution is not in the scope of this article, we emphasize that in our calculations there are candidates for GRBs and high spin BHs formation and there are candidates for collapse without a GRB, at least a part of systems can possess a low value of the BH spin.

Future statistics of gravitational wave events and follow-up observations of such events in electromagnetic spectra probably will be able to clarify the viability of different suggested mechanisms of GW formation \citep{abbott2016b,breivik2016,rodriguez2016a}. The existence of such channels as suggested by e. g. \citet{sigurdsson1993,liu2015,antonini2016,blinnikov2016,
inayoshi2016,loeb2016,mandel2016,mapelli2016,mink2016}, and \citet{rodriguez2016a} probably will be proved, and even such unusual applications as made by \citet{gorkavyi2016} about a repulsive force that probably can explain the expansion and the acceleration of the expansion of the Universe also can become testable. Constraints on the evolutionary scenario made by \citet{kushnir2016}, \citet{belczynski2016b} and in the present study explain the evolutionary path of the GW150914 gravitational radiation burst. We can find that the classical prediction of GW origin as a merger of two remnants of a binary star evolution is in adequate agreement with observed properties of GW150914.

The weak stellar wind may be a consequence of the low abundance of heavy elements, see e. g. \citet{massevich1988}, however the ``Scenario Machine'' program does not take into account metallicity dependences. We can imitate metallicity differences of the stellar wind strength by changing coefficients $\alpha$ and $\alpha_{WR}$. So, our results could be useful not only for applications of the evolution of binaries to gravitational wave observations, but also for studies of WR stars evolution like e. g. \citep{mcclelland2016}. A zero or very low kick velocity is a feature of massive merging BHs (with both BH masses $\ge 25 M_{\odot}$) in our calculations using our population synthesis code, and this statement should not be expanded to other ranges of BH masses or to pairs that include a neutron star. In addition to other studies we suggest a new evolutionary track (see Table \ref{track2}) that can lead to the formation of merging binary BHs. The masses of the black holes discovered by LIGO GW150914 highly exceed the masses of black holes candidates in X-ray novae systems, see, e. g., BH mass functions calculated by \citet{fryer2001,bogomazov2005,kochanek2015}. A probable explanation of this fact may be that if the initial mass ratio is low $q=M_2/M_1\sim 0.01$, the smaller star is not able to get to the main sequence and it is evaporated by the blue star of which the luminosity exceeds the luminosity of the smaller companion by a factor of some $\sim 10^6$. Therefore massive BHs can not be found in X-ray novae binaries with low-mass donor stars \citep{lipunov2016}. Also a survived low mass star can be completely consumed by a massive companion by spiral in it during CE stage. These facts eliminate the possibility of massive BHs merging with white dwarfs, and this statement is testable by LIGO. The relatively high BH masses in GW150914 are in a good agreement with parameters of evolution of massive stars and in accord with selection effects and the detectability of mergers by LIGO.

We conclude that GW150914 does not put final constraints on the evolutionary scenario parameters; some of them, for example, $\alpha_{CE}$ remains very uncertain. \citet{massevich1988} found that the masses of collapsed cores of massive stars (with masses $\ge 30 M_{\odot}$) after carbon and oxygen burning are given by:

\begin{equation}
\frac{M_{BH}}{M_{\odot}}\approx 0.05 \left(\frac{M_1}{M_{\odot}}\right)^{1.4},
\label{core-tutukov}
\end{equation}

\noindent where $M_{BH}$ is the mass of the BH formed under the assumption that all matter of the collapsing core after C and O buring falls into the BH\footnote{The nuclear energy of C+O burning is less than the gravitational energy.}, $M_1$ is the initial mass of the MS star. Our set of parameters (d) (see Section \ref{results}) seems to produce a binary with masses of two BHs that are very similar to the masses measured in GW150914. In the limits of modern uncertainties  of crucial evolutionary parameters, GW150914 is an example of one of the most massive binary stars.

Our sets of parameters (a), (b), (c), and (e) are expected to be useful as the theoretical and observational basis of knowledge about the evolution of binaries is still incomplete. The evolution of very massive stars can be limited by intense stellar wind and potentially goes through somewhat different ways than descibed here. In all currently published LIGO events the BH masses are $\lesssim 40 M_{\odot}$, so our set of parameters (d) seems to be preferable from the observational point of view. Future work of LIGO probably will be able to narrow uncertainties in mass loss rate by optical stars, in the CE efficiency and in masses of SN remnants.

\section*{Ackhowledgments}

The work was supported by the Russian Foundaton Basic Research (projects 14-02-00825, 15-02-04053), by a grant for leading scientific schools \\ NSh-9670.2016.2, by the Russian Science Foundation (project 16-12-00085).

\clearpage

\begin{table}
\centering
\begin{minipage}{120mm}
\caption{Evolutionary track of merging BHs calculated using the following set of parameters: $\alpha=0.1$, $\alpha_{WR}=0.1$, $\alpha_{CE}=0.7$, $k_{bh}=1$, $v_0=0$. Here ``System'' depicts evolutionary status of the stars in the binary, $\Delta T$ is the duration of an evolutionary stage, $M_1$ is the mass of the primary (initially more massive) star, $M_2$ is the mass of the secondary star, $a$ is the major semi-axis, $e$ is the eccentricity of the binary, $T$ is the time since the formation of the binary. $\Delta T$ and $T$ are in units of $10^6$ years, $M_1$ and $M_2$ are in solar masses, $a$ is in solar radii. Notations of system's state are the following (see \citep{lipunov2009} for details): ``I'' is a main sequence star, ``3'' is a star filling its Roche lobe, ``3E'' is a star that fills its Roche lobe on an evolutionary time scale, ``WR'' is a Wolf-Rayet star, ``BH'' is a black hole, ``SH'' is a black hole with a super critical accretion rate. The amount of energy that is radiated in gravitational waves is not taken into account in the last stage (a single BH). }
\label{track1}
\begin{tabular}{@{}ccccccc@{}}
\hline
System & $\Delta T$ & $M_1$ & $M_2$ & $a$ & $e$ & $T$ \\
\hline
I+I & 2.5 & 88.92 & 38.41 & 39 & 0 & 0 \\
I+I & & 87.44 & 37.41 & 40 & 0 & 2.5 \\
3+I & $3.7\cdot 10^{-3}$ & 87.44 & 37.41 & 40 & 0 & 2.5 \\
3+I & & 62.43 & 62.43 & 28 & 0 & 2.5 \\
3E+I & 0.63 & 62.43 & 62.43 & 28 & 0 & 2.5 \\
3E+I & & 53.53 & 68.24 & 30 & 0 & 3.1 \\
WR+I & 0.2 & 53.53 & 68.24 & 30 & 0 & 3.1 \\
WR+I & & 48.18 & 68.18 & 31 & 0 & 3.3 \\
\multicolumn{6}{c}{Supenova type Ib, explosion 1 in Table \ref{kerr}} \\
BH+I & $2.5\cdot 10^{-2}$ & 38.54 & 68.18 & 34 & 0.09 & 3.3 \\
BH+I & & 38.54 & 68.17 & 34 & 0 & 3.4 \\
SH+3 & $8.3\cdot 10^{-3}$ & 38.54 & 68.17 & 34 & 0 & 3.4 \\
SH+3 & & 38.55 & 38.55 & 34 & 0 & 3.4 \\
SH+3E & 0.66 & 38.55 & 38.55 & 34 & 0 & 3.4 \\
SH+3E & & 38.71 & 36.96 & 35 & 0 & 4.0 \\
BH+WR & 0.24 & 38.71 & 36.96 & 35 & 0 & 4.0 \\
BH+WR & & 38.71 & 33.26 & 36 & 0 & 4.3 \\
\multicolumn{6}{c}{Supenova type Ib, explosion 2 in Table \ref{kerr}} \\
BH+BH & $5.1\cdot 10^3$ & 38.71 & 26.61 & 41 & 0.10 & 4.3 \\
\multicolumn{6}{c}{Coalescence} \\
BH & & 65.32 & & & & $5.1\cdot 10^3$ \\
\hline
\end{tabular}
\end{minipage}
\end{table}

\clearpage

\begin{table}
\centering
\begin{minipage}{120mm}
\caption{Evolutionary track of merging BHs calculated using the following set of parameters: $\alpha=0.1$, $\alpha_{WR}=0.1$, $\alpha_{CE}=0.7$, $k_{bh}=0.8$, $v_0=0$. ``CE'' is the common envelope stage, other notations and units in this Table are the same as in the Table \ref{track1}.}
\label{track2}
\begin{tabular}{@{}ccccccc@{}}
\hline
System & $\Delta T$ & $M_1$ & $M_2$ & $a$ & $e$ & $T$ \\
\hline
I+I & 2.6 & 80.47 & 24.76 & 180 & 0 & 0 \\
I+I &  & 79.30 & 24.47 & 180 & 0 & 2.6 \\
3+I, CE & 0.01 & 79.30 & 24.47 & 180 & 0 & 2.6 \\
3+I, CE &  & 46.55 & 43.90 & 12 & 0 & 2.6 \\
WR+3E & 0.21 & 46.55 & 43.90 & 12 & 0 & 2.6 \\
WR+3E & & 49.10 & 37.31 & 13 & 0 & 2.8 \\
\multicolumn{6}{c}{Supenova type Ib, explosion 3 in Table \ref{kerr}} \\
BH+3E & 0.56 & 39.28 & 37.31 & 15 & 0.13 & 2.8 \\
BH+3E & & 39.28 & 37.31 & 13 & 0 & 3.3 \\
BH+WR & 0.34 & 39.28 & 37.31 & 13 & 0 & 3.3 \\
BH+WR & & 39.28 & 33.58 & 14 & 0 & 3.7 \\
\multicolumn{6}{c}{Supenova type Ib, explosion 4 in Table \ref{kerr}} \\
BH+BH & 110 & 39.28 & 26.86 & 16 & 0.10 & 3.7 \\
\multicolumn{6}{c}{Coalescence} \\
BH & & 66.14 & & & & 110 \\
\hline
\end{tabular}
\end{minipage}
\end{table}

\clearpage

\begin{table}
\centering
\begin{minipage}{120mm}
\caption{Evolutionary track of merging BHs calculated using the following set of parameters: $\alpha=0.1$, $\alpha_{WR}=0.1$, $\alpha_{CE}=0.7$, $k_{bh}=0.55$, $v_0=0$. ``CE'' is the common envelope stage, other notations and units in this Table are the same as in the Tables \ref{track1} and \ref{track2}.}
\label{track3}
\begin{tabular}{@{}ccccccc@{}}
\hline
System & $\Delta T$ & $M_1$ & $M_2$ & $a$ & $e$ & $T$ \\
\hline
I+I & 2.4 & 117.59 & 79.59 & 140 & 0 & 0 \\
I+I & & 114.69 & 78.51 & 150 & 0 & 2.4 \\
3+I & $8.1\cdot 10^{-4}$ & 114.69 & 78.51 & 150 & 0 & 2.4 \\
3+I & & 85.88  & 85.88 & 140 & 0 & 2.4 \\
3E+I & 0.61 & 85.88 & 85.88 & 140 & 0 & 2.4 \\
3E+I & & 79.17 & 87.07 & 150 & 0 & 3.0 \\
WR+I & 0.16 & 79.17 & 87.07 & 150 & 0 & 3.0 \\
WR+I & & 71.25 & 86.98 & 160 & 0 & 3.2 \\
\multicolumn{6}{c}{Supenova type Ib, explosion 5 in Table \ref{kerr}} \\
BH+I & 5.7E-03 & 39.19 & 86.98 & 210 & 0.26 & 3.2 \\
BH+I & & 39.19 & 86.98 & 210 & 0.26 & 3.2 \\
SH+3S & $2.9\cdot 10^{-4}$ & 39.19 & 86.98 & 210 & 0.26 & 3.2 \\
SH+3S & & 39.19 & 82.06 & 190 & 0.23 & 3.2 \\
SH+3, CE & 0.01 & 39.19 & 82.06 & 190 & 0.23 & 3.2 \\
SH+3, CE & & 39.19 & 51.98 & 16 & 0 & 3.2 \\
BH+WR & 0.2 & 39.19 & 51.98 & 16 & 0 & 3.2 \\
BH+WR & & 39.19 & 46.78 & 17 & 0 & 3.4 \\
\multicolumn{6}{c}{Supenova type Ib, explosion 6 in Table \ref{kerr}} \\
BH+BH & 580 & 39.19 & 25.73 & 26 & 0.33 & 3.4 \\
\multicolumn{6}{c}{Coalescence} \\
BH & & 64.92 & & & & 580 \\
\hline
\end{tabular}
\end{minipage}
\end{table}

\clearpage

\begin{table}
\centering
\begin{minipage}{120mm}
\caption{An example of an evolutionary track of massive close binaries that produces merging BHs calculated with following parameters: $\alpha=0.1$, $\alpha_{WR}=0.1$, $\alpha_{CE}=0.7$, $k_{bh}=1$, $v_0=0$. Notations and units are the same as in the Table \ref{track1}.}
\label{track4}
\begin{tabular}{@{}ccccccc@{}}
\hline
System & $\Delta T$ & $M_1$ & $M_2$ & $a$ & $e$ & $T$ \\
\hline
I+I & 3.2 & 76.33 & 57.21 & 210 & 0 & 0 \\
I+I & & 75.03 & 56.59 & 210 & 0 & 3.2 \\
II+I & 0.26 & 75.03 & 56.59 & 210 & 0 & 3.2 \\
II+I & & 64.63 & 56.54 & 230 & 0 & 3.5 \\
II+II & $6.3\cdot 10^{-2}$ & 64.63 & 56.54 & 230 & 0 & 3.5 \\
II+II & & 63.15 & 55.64 & 240 & 0 & 3.5 \\
3+II, CE & $1\cdot 10^{-2}$ & 63.15 & 55.64 & 240 & 0 & 3.5 \\
3+II, CE & & 43.23 & 28.87 & 45 & 0 & 3.6 \\
WR+WR & 0.22 & 43.23 & 28.87 & 45 & 0 & 3.6 \\
WR+WR & & 38.90 & 26.57 & 49 & 0 & 3.8 \\
\multicolumn{6}{c}{Supenova Ib, explosion 7 in Table \ref{kerr}} \\
BH+WR & $5.7\cdot 10^{-2}$ & 38.90 & 26.57 & 49 & 0 & 3.8 \\
BH+WR & & 38.90 & 26.03 & 50 & 0 & 3.8 \\
\multicolumn{6}{c}{Supenova Ib, explosion 8 in Table \ref{kerr}} \\
BH+BH & $1.2\cdot 10^{4}$ & 38.90 & 26.03 & 50 & 0 & $1.2\cdot 10^{4}$ \\
\multicolumn{6}{c}{Coalescence} \\
BH & & 64.94 & & & & $1.2\cdot 10^{4}$ \\
\hline
\end{tabular}
\end{minipage}
\end{table}

\clearpage

\begin{table}
\centering
\begin{minipage}{120mm}
\caption{Orbital periods before supernova explosions and dimensionless Kerr parameters of collapsing cores of WR stars. Notations in the Table are the following. In the first column ``1'' is the first explosion in the track in Table \ref{track1}, ``2'' is the second explosion in the track in Table \ref{track1}, ``3'' is the first explosion in the track in Table \ref{track2}, ``4'' is the second explosion in the track in Table \ref{track2}, ``5'' is the first explosion in the track in Table \ref{track3}, ``6'' is the second explosion in the track in Table \ref{track3}, ``7'' is the first explosion in the track in Table \ref{track4}, ``8'' is the second explosion in the track in Table \ref{track4}. $P_{orb}$ is the orbital period of the system just before the explosion, $a_{K1}$ is the dimensionless Kerr parameter for $k_I=0.1$, and $a_{K2}$ is the dimensionless Kerr parameter for $k_I=0.4$. $a_K\ge 1$ means that the core can't collapse into BH immediately and a long gamma ray burst is able to be produced before the matter disappers under the event horizon.}
\label{kerr}
\begin{tabular}{@{}cccc@{}}
\hline
Explosion & $P_{orb}$, days & $a_{K1}$ & $a_{K2}$ \\
\hline
1 & 1.6 & 2.2 & 8 \\
2 & 2.5 & 1.3 & 5.3 \\
3 & 0.5 & 7 & 28 \\
4 & 0.6 & 4.3 & 17.1 \\
5 & 16.1 & 0.33 & 1.4 \\
6 & 0.76 & 5 & 20 \\
7 & 4.2 & 0.66 & 2.6 \\
8 & 4.4 & 0.29 & 1.14 \\
\hline
\end{tabular}
\end{minipage}
\end{table}

\clearpage

\clearpage

\clearpage

\clearpage

\clearpage

\clearpage

\clearpage

\clearpage

\clearpage

\clearpage

\clearpage

\clearpage

\clearpage

\clearpage

\clearpage

\clearpage

\clearpage

\clearpage

\clearpage

\clearpage

\clearpage

\clearpage

\clearpage

\clearpage

\begin{thebibliography}{00}

\bibitem[\protect\citeauthoryear{Abbott et al.}{2016a}]{abbott2016a} Abbott B. P., et al. (LIGO Scientific Collaboration and Virgo Collaboration), 2016a, Physical Review Letters, 116, 061102

\bibitem[\protect\citeauthoryear{Abbott et al.}{2016b}]{abbott2016b} Abbott B. P., et al. (LIGO Scientific Collaboration and Virgo Collaboration), 2016b, Living Reviews in Relativity, 19, 1

\bibitem[\protect\citeauthoryear{Abbott et al.}{2016c}]{abbott2016c} Abbott B. P., et al. (LIGO Scientific Collaboration and Virgo Collaboration), 2016c, ApJL, 818, L22

\bibitem[\protect\citeauthoryear{Abbott et al.}{2016d}]{abbott2016d} Abbott B. P. , et al. (LIGO Scientific Collaboration and Virgo Collaboration), 2016d, Physical Review Letters, 116, 241103

\bibitem[\protect\citeauthoryear{Abbott et al.}{2016e}]{ligo2016} Abbott B. P. et al., 2016, Physical Review Letters, 116, id.241102

\bibitem[\protect\citeauthoryear{Abubekerov et al.}{2009}]{abubekerov2009} Abubekerov M. K., Antokhina E. A., Bogomazov A. I., Cherepashchuk A. M., 2009, Astronomy Reports, 53, 232

\bibitem[\protect\citeauthoryear{Antonini et al.}{2016}]{antonini2016} Antonini F., Chatterjee S., Rodriguez C., Morscher M., Pattabiraman B., Kalogera V., Rasio F., 2016, ApJ, 816, id. 65

\bibitem[\protect\citeauthoryear{Bauer \& Brandt}{2009}]{bauer2004} Bauer F. E., Brandt W. N., 2004, ApJL, 601, L67

\bibitem[\protect\citeauthoryear{Belczynski et al.}{2016a}]{belczynski2016} Belczynski K. et al., 2016a, ApJ, 819, 108

\bibitem[\protect\citeauthoryear{Belczynski et al.}{2016b}]{belczynski2016b} Belczynski K, Holz D. E., Bulik T., O'Shaughnessy R., 2016b, Nature, 534, 512

\bibitem[\protect\citeauthoryear{Belczynski et al.}{2016c}]{belczynski2016c} Belczynski K., Heger A., Gladysz W., Ruiter A., Woosley S., Wiktorowicz G., Chen H.-Y., Bulik T., O'Shaughnessy R., Holz D., Fryer C., Berti E., 2016, Astronomy and Astrophysics, 594, id. A97

\bibitem[\protect\citeauthoryear{Benaglia et al.}{2005}]{benaglia2005} Benaglia P., Romero G. E., Koribalski B., Pollock A. M. T., 2005, A\&A, 440, 743

\bibitem[\protect\citeauthoryear{Binder et al.}{2015}]{binder2015} Binder B., Gross J., Williams B. F.,  Simons D., 2015, MNRAS, 451, 4471

\bibitem[\protect\citeauthoryear{Blinnikov et al.}{2016}]{blinnikov2016} Blinnikov S., Dolgov A., Porayko N. K., Postnov K., 2016, Journal of Cosmology and Astroparticle Physics, 11, id. 036

\bibitem[\protect\citeauthoryear{Bogomazov, Abubekerov \& Lipunov}{2005}]{bogomazov2005} Bogomazov A. I., Abubekerov M. K., Lipunov V. M., 2005, Astronomy Reports, 49, 644
	
\bibitem[\protect\citeauthoryear{Bogomazov, Lipunov \& Tutukov}{2007}]{bogomazov2007} Bogomazov A. I., Lipunov V. M., Tutukov A. V., 2007, 	Astronomy Reports, 51, 308	
	
\bibitem[\protect\citeauthoryear{Bogomazov, Lipunov \& Tutukov}{2008}]{bogomazov2008a}	Bogomazov A. I., Lipunov V. M., Tutukov A. V., 2008, Astronomy Reports, 52, 463

\bibitem[\protect\citeauthoryear{Bogomazov \& Cherepashchuk}{2008}]{bogomazov2008b} Bogomazov A. I., Cherepashchuk A. M., 2008, Astronomy Reports, 52, 1009

\bibitem[\protect\citeauthoryear{Bogomazov}{2014}]{bogomazov2014} Bogomazov A. I., 2014, Astronomy Reports, 58, 126

\bibitem[\protect\citeauthoryear{Bonanos et al.}{2004}]{bonanos2004} Bonanos A. Z., 2004, ApJ, 611, L33

\bibitem[\protect\citeauthoryear{Brandt et al.}{1997}]{brandt1997} Brandt W. N., Ward M. J., Fabian A. C., Hodge P. W., 1997, MNRAS, 291, 709

\bibitem[\protect\citeauthoryear{Breivik et al.}{2016}]{breivik2016} Breivik K., Rodriguez C. L., Larson S. L., Kalogera V., Rasio F. A., 2016, ApJL, 830, article id. L18

\bibitem[\protect\citeauthoryear{Caprano et al.}{2007}]{caprano2007} Carpano S. et al., 2007, A\&A, 466, L17

\bibitem[\protect\citeauthoryear{Caraveo et al.}{1989}]{caraveo1989} Caraveo P. A., Bignami G. F., Goldwurm A. et al., 1989, ApJ, 338, 338

\bibitem[\protect\citeauthoryear{Cherepashchuk}{1976}]{cherepashchuk1976} Cherepashchuk A. M., 1976, Soviet Astronomy Letters, 2, 138

\bibitem[\protect\citeauthoryear{Cherepashchuk}{1981}]{cherepashchuk1981} Cherepashchuk A. M., 1981, MNRAS, 194, 761

\bibitem[\protect\citeauthoryear{Cherepashchuk et al.}{1984}]{cherepashchuk1984} Cherepashchuk A. M., Eaton J. A., Khaliullin Kh. F., 1984, ApJ, 281, 774

\bibitem[\protect\citeauthoryear{Cherepashchuk}{1990}]{cherepashchuk1990} Cherepashchuk A. M., 1990, Soviet Astronomy, 34, 481

\bibitem[\protect\citeauthoryear{Cherepashchuk}{1991}]{cherepashchuk1991} Cherepashchuk A. M., 1991, Wolf-Rayet Stars and Interrelations with Other Massive Stars in Galaxies: Proceedings of the 143rd Symposium of the International Astronomical Union, Edited by Karel A. van der Hucht and Bambang Hidayat, p. 187	

\bibitem[\protect\citeauthoryear{Cherepashchuk}{2001}]{cherepashchuk2001} Cherepashchuk A. M., 2001, Astronomy Reports, 45, 120

\bibitem[\protect\citeauthoryear{Cherepashchuk}{2003}]{cherepashchuk2003} Cherepashchuk A. M., 2003, Physics Uspekhi, 46, 335

\bibitem[\protect\citeauthoryear{Cherepashchuk}{2013}]{cherepashchuk2013} Cherepashchuk A. M., 2013, close binary stars, Fizmat-lit, Moscow, 176, 462 (in Russian)

\bibitem[\protect\citeauthoryear{Cherepashchuk et al.}{2013}]{cherepashchuk2013b} Cherepashchuk A. M., Sunyaev R. A., Molkov S. V., Antokhina E. A., Postnov K. A., Bogomazov A. I., 2013, MNRAS, 436, 2004

\bibitem[\protect\citeauthoryear{Crowther \& Dessart}{1998}]{crowther1998} Crowther P. A., Dessart L., 1998, MNRAS, 296, 622

\bibitem[\protect\citeauthoryear{Crowther et al.}{2003}]{crowther2003} Crowther P. A. et al., 2003, A\&A, 404, 483

\bibitem[\protect\citeauthoryear{Crowther et al.}{2010}]{crowther2010} Crowther P. et al., 2010, MNRAS, 403, L41

\bibitem[\protect\citeauthoryear{de Mink \& Mandel}{2016}]{mink2016} de Mink S. E., Mandel I., 2016, MNRAS, 460, 3545

\bibitem[\protect\citeauthoryear{de Koter et al.}{1997}]{koter1997} de Koter A., Heap S. R., Hubeny I., 1997, ApJ, 477, 792

\bibitem[\protect\citeauthoryear{Drissen et al.}{1995}]{drissen1995} Drissen L., Moffat A. F. J., Walborn N. R., Shara M. M., 1995, AJ, 110, 2235

\bibitem[\protect\citeauthoryear{Fabrika}{2004}]{fabrika2004} Fabrika S., 2004, Astrophys. Space Phys. Rev., 12, 1

\bibitem[\protect\citeauthoryear{Fryer \& Kalogera}{2001}]{fryer2001} Fryer C., Kalogera V., 2001, ApJ, 554, 548

\bibitem[\protect\citeauthoryear{Fryer, Woosley \& Heger}{2001}]{fryer2001b} Fryer C., Woosley S, Heger A., 2001, ApJ, 550, 372

\bibitem[\protect\citeauthoryear{Fryer et al.}{2012}]{fryer2012} Fryer C. L., Belczynski K., Wiktorowicz G., Dominik M., Kalogera V., Holz D. E., 2012, ApJ, 749, id. 91

\bibitem[\protect\citeauthoryear{Gammie, Shapiro \& McKinney}{2004}]{gammie2004} Gammie C. F., Shapiro S. L., McKinney J. C., 2004, ApJ, 602, 312

\bibitem[\protect\citeauthoryear{Gorkavyi \& Vasilkov}{2016}]{gorkavyi2016} Gorkavyi N., Vasilkov A., 2016, MNRAS, 461, 2929

\bibitem[\protect\citeauthoryear{Hillier}{1991}]{hillier1991}	 Hillier D. J., 1991, Astronomy and Astrophysics, 247, 455

\bibitem[\protect\citeauthoryear{Esposito et al.}{2013}]{esposito2013} Esposito P., Israel G. L., Sidoli L., Mapelli M., Zampieri L., Motta S. E., 2013, MNRAS, 436, 3380

\bibitem[\protect\citeauthoryear{Hanson, Still \& Fender}{2000}]{hanson2000} Hanson M. M. , Still M. D., Fender R. P., 2000, ApJ, 541, 308

\bibitem[\protect\citeauthoryear{Heger \& Woosley}{2002}]{heger2002} Heger A., Woosley S., 2002, ApJ, 567, 532

\bibitem[\protect\citeauthoryear{Heger et al.}{2003}]{heger2003} Heger A., Fryer C. L., Woosley S. E., Langer N., Hartmann D. H., 2003, ApJ, 591, 288

\bibitem[\protect\citeauthoryear{Hillier}{2003}]{hillier2003} Hillier D. J., 2003, Proceedings of IAU Symposium 212, eds. K. van der Hucht, A. Herrero,  E. Cesar, San Francisco: Astronomical Society of the Pacific, 70

\bibitem[\protect\citeauthoryear{Howarth \& Schmutz}{1992}]{howarth1992} Howarth I. D., Schmutz W., 1992, Astronomy and Astrophysics, 261, 503

\bibitem[\protect\citeauthoryear{Inayoshi et al.}{2016}]{inayoshi2016} Inayoshi K., Kashiyama K., Visbal E., Haiman Z., 2016, MNRAS, 461, 2722

\bibitem[\protect\citeauthoryear{Kochanek}{2015}]{kochanek2015}	Kochanek C. S., 2015, MNRAS, 446, 1213

\bibitem[\protect\citeauthoryear{Kornilov \& Lipunov}{1983}]{kornilov1983} Kornilov V. G., Lipunov V. M., 1983, Soviet Astronomy, 27, 334

\bibitem[\protect\citeauthoryear{Kushnir et al.}{2016}]{kushnir2016} Kushnir D., Zaldarriaga M., Kollmeier J., Waldman R., 2016, MNRAS, 462, 844

\bibitem[\protect\citeauthoryear{Langer}{1989a}]{langer1989a} Langer N., 1989a, Astronomy and Astrophysics, 210, 93
	
\bibitem[\protect\citeauthoryear{Langer}{1989b}]{langer1989b} Langer, N., 1989b, Astronomy and Astrophysics, 220, 135

\bibitem[\protect\citeauthoryear{Laycock, Cappallo \& Moro}{2015}]{laycock2015a} Laycock S. G. T., Cappallo R. C., Moro M. J., 2015, MNRAS, 446, 1399

\bibitem[\protect\citeauthoryear{Laycock, Maccarone \& Christodoulou}{2015b}]{laycock2015b} Laycock S. G. T., Maccarone T. J., Christodoulou D. M., 2015, MNRAS, 452, L31

\bibitem[\protect\citeauthoryear{Laycock et al.}{2017}]{laycock2017} Laycock S., Christodoulou D., Williams B., Binder B., Prestwich A., 2017, ApJ, 836, id. 51

\bibitem[\protect\citeauthoryear{Lipunov}{1982}]{lipunov1982} Lipunov V. M., 1982, Soviet Astronomy Letters, 8, 194

\bibitem[\protect\citeauthoryear{Lipunov, Postnov \& Prokhorov}{1996}]{lipunov1996} Lipunov V. M., Postnov K. A., Prokhorov M. E., 1996, Astrophys. Space Phys. Rev., 9, 1

\bibitem[\protect\citeauthoryear{Lipunov, Postnov \& Prokhorov}{1997}]{lipunov1997a} Lipunov V. M., Postnov K. A., Prokhorov M. E., 1997, New Astronomy, 2, 43

\bibitem[\protect\citeauthoryear{Lipunov et al.}{2009}]{lipunov2009} Lipunov V. M., Postnov K. A., Prokhorov M. E., Bogomazov A. I., 2009,  Astronomy Reports, 53, 915

\bibitem[\protect\citeauthoryear{Lipunov et al.}{2016}]{lipunov2016} Lipunov V. M., Kornilov V. G., Gorbovskoy E. S., Tiurina N. A., Balanutsa P., Kuznetsov A., 2016, New Astronomy, 51, 122

\bibitem[\protect\citeauthoryear{Lipunova et al.}{2009}]{lipunova2009} Lipunova G. V., Gorbovskoy E. S., Bogomazov A. I., Lipunov V. M., 2009, MNRAS, 397, 1695

\bibitem[\protect\citeauthoryear{Liu, Lai \& Yuan}{2015}]{liu2015} Liu B., Lai D., Yuan Y.-F., 2015, Physical Review D, 92, id. 124048 

\bibitem[\protect\citeauthoryear{Loeb}{2016}]{loeb2016} Loeb A., 2016, ApJL, 819, L21

\bibitem[\protect\citeauthoryear{Long et al.}{1981}]{long1981} Long K. S., Dodorico S., Charles P. A., Dopita M. A., 1981, ApJL, 246, L61

\bibitem[\protect\citeauthoryear{Lozinskaya \& Moiseev}{2007}]{lozinskaya2007} Lozinskaya T. A., Moiseev A. V., 2007, MNRAS, 381, L26

\bibitem[\protect\citeauthoryear{Lozinskaya et al.}{2008}]{lozinskaya2008} Lozinskaya T. A., Moiseev A. V., Podorvanyuk N. Yu., Burenkov A. N., 2008, Astronomy Letters, 34, 217

\bibitem[\protect\citeauthoryear{Mandel \& de Mink}{2016}]{mandel2016} Mandel I., de Mink S. E., 2016, 	MNRAS, 458, 2634

\bibitem[\protect\citeauthoryear{Mapelli}{2016}]{mapelli2016} Mapelli M., 2016, MNRAS, 459, 3432

\bibitem[\protect\citeauthoryear{Masevich et al.}{1979}]{massevich1979} Masevitch A. G., Popova E. I., Tutukov A. V., Iungelson L. R., 1979, Astrophysics and Space Science, 62, 451

\bibitem[\protect\citeauthoryear{Masevich \& Tutukov}{1988}]{massevich1988} Masevich A. G., Tutukov A. V., 1988, Stellar Evolution: Theory and Observations, Moscow: Nauka (in Russian)

\bibitem[\protect\citeauthoryear{Mereghetti et al.}{1994}]{mereghetti1994} Mereghetti S., Belloni T., Shara M., Drissen L., 1994, ApJ, 424, 943

\bibitem[\protect\citeauthoryear{McClelland \& Eldridge}{2016}]{mcclelland2016}	McClelland L. A. S., Eldridge J. J., 2016, MNRAS, 459, 1505

\bibitem[\protect\citeauthoryear{Moffat \& Niemela}{1984}]{moffat1984} Moffat A. F. J., Niemela V. S.,
1984, ApJ, 284, 631

\bibitem[\protect\citeauthoryear{Moffat et al.}{1985}]{moffat1985} Moffat A. F. J., Seggewiss W., Shara M. M., 1985, ApJ, 295, 109

\bibitem[\protect\citeauthoryear{Moffat et al.}{1988}]{moffat1988} Moffat A. F.J., Drissen L., Lamontagne R., Robert C., 1988, ApJ, 334, 1038

\bibitem[\protect\citeauthoryear{Moffat et al.}{2004}]{moffat2004} Moffat A. F. J., Poitras V., Marchenko S. V. et al., 2004, AJ, 128, 2854

\bibitem[\protect\citeauthoryear{Nazin \& Postnov}{1995}]{nazin1995} Nazin S. N., Postnov K. A., 1995, A\&A, 303, 789

\bibitem[\protect\citeauthoryear{Niemela et al.}{2008}]{niemela2008} Niemela V. S., Gamen R. C., Barba R. H. et al, 2008, MNRAS, 389, 1447

\bibitem[\protect\citeauthoryear{Orosz et al.}{2007}]{orosz2007} Orosz J. A. et al., 2007, Nature, 449, 872

\bibitem[\protect\citeauthoryear{Pavlovskii et al.}{2017}]{pavlovskii2017} Pavlovskii K., Ivanova N., Belczynski K., Van K. X., 2017, MNRAS, 465, 2092

\bibitem[\protect\citeauthoryear{Peres et al.}{1989}]{peres1989} Peres G., Reale F., Collura A., Fabbiano G., 1989, ApJ, 336, 140

\bibitem[\protect\citeauthoryear{Pietsch et al.}{2004}]{pietsch2004} Pietsch W. et al., 2004, A\&A, 413, 879

\bibitem[\protect\citeauthoryear{Pietsch et al.}{2006}]{pietsch2006} Pietsch W. et al., 2006, ApJ, 646, 420

\bibitem[\protect\citeauthoryear{Popescu \& Hanson}{2014}]{popescu2014} Popescu B., Hanson M., 2014,	
ApJ, 780, id. 27

\bibitem[\protect\citeauthoryear{Popov \& Prokhorov}{2007}]{popov2007} Popov S. B., Prokhorov M. E., 2007, Physics Uspekhi, 50, 1123

\bibitem[\protect\citeauthoryear{Prestwich et al.}{2007}]{prestwich2007} Prestwich A. H., Kilgard R., Crowther P. A. et al., 2007, ApJ, 669, L21
	
\bibitem[\protect\citeauthoryear{Prinja et al.}{1990}]{prinja1990} Prinja R. K., Barlow M. J., Howarth I. D., 1990, ApJ, 361, 607	
	
\bibitem[\protect\citeauthoryear{Prinja et al.}{1991}]{prinja1991} Prinja R. K., Barlow M. J., Howarth I. D., 1991, ApJ, 383, 466

\bibitem[\protect\citeauthoryear{Puls et al.}{2003}]{puls2003} Puls J. et al., 2003, Proceedings of IAU Symposium 212, eds K. van der Hucht, A. Herrero, E. Cesar, San Francisco: Astronomical Society of the Pacific, 61

\bibitem[\protect\citeauthoryear{Rauw et al.}{1996}]{rauw1996} Rauw G., Vreux G.-M., Gosset E., et al., 1996, Astronomy and Astrophysics, 306, 771

\bibitem[\protect\citeauthoryear{Rauw et al.}{2004}]{rauw2004} Rauw G., De Becker M., Naze Y., et al., 2004, Astronomy and Astrophysics, 420, L9

\bibitem[\protect\citeauthoryear{Rodriguez et al.}{2016a}]{rodriguez2016a} Rodriguez C. L., Haster C.-J., Chatterjee S., Kalogera V., Rasio F. A., 2016, ApJL, 824, L8

\bibitem[\protect\citeauthoryear{Rodriguez et al.}{2016b}]{rodriguez2016b} Rodriguez C. L., Zevin M., Pankow C., Kalogera V., Rasio F. A., 2016, ApJL, 832, L2

\bibitem[\protect\citeauthoryear{Sana et al.}{2012}]{sana2012} Sana H., de Mink S. E., de Koter A., Langer N., Evans C. J., Gieles M., Gosset E., Izzard R. G., Le Bouquin J.-B., Schneider F. R. N., 2012, Science, 337, 444

\bibitem[\protect\citeauthoryear{Schmutz, Geballe \& Schild}{1996}]{schmutz1996} Schmutz W., Geballe T. R., Schild H., 1996, Astronomy and Astrophysics, 311, L25

\bibitem[\protect\citeauthoryear{Schnurr et al.}{2008}]{schnurr2008} Schnurr O., Casoli J., Chene A.-N. et al., 2008, MNRAS, 389, L38

\bibitem[\protect\citeauthoryear{Schnurr et al.}{2009}]{schnurr2009} Schnurr, O., Moffat A. F. J., Villar-Sbaffi A. et al., 2009, MNRAS, 395, 823

\bibitem[\protect\citeauthoryear{Schweickhardt et al.}{1999}]{schweickhardt1999} Schweickhardt J., Schmutz W., Stahl O., et al., 1999, Astronomy and Astrophysics, 347, 127

\bibitem[\protect\citeauthoryear{Shara et al.}{1991}]{shara1991} Shara M. M., Moffat A. F. J., Smith L. F., Potter M., 1991, AJ, 102, 716

\bibitem[\protect\citeauthoryear{Sigurdsson \& Hernquist}{1993}]{sigurdsson1993} Sigurdsson S., Hernquist L., 1993, Nature, 364, 423

\bibitem[\protect\citeauthoryear{Silverman \& Filippenko}{2008}]{silverman2008} Silverman J. M., Filippenko A. V., 2008, ApJ, 678, L17

\bibitem[\protect\citeauthoryear{Stark \& Saia}{2003}]{stark2003} Stark M. J., Saia M., 2003, ApJL, 587,
L101

\bibitem[\protect\citeauthoryear{Steiner et al.}{2016}]{steiner2016} Steiner J. F. et al., 2016, ApJ, 817, 154

\bibitem[\protect\citeauthoryear{St.-Louis et al.}{1988}]{stlouis1988} St.-Louis N., Moffat A. F. J., Drissen L., Bastien P., Robert C., 1988, ApJ, 330, 286

\bibitem[\protect\citeauthoryear{Sukhbold et al.}{2016}]{sukhbold2016} Sukhbold T., Ertl T., Woosley S. E., Brown J. M.. Janka H.-T., 2016, ApJ, 821, id. 38

\bibitem[\protect\citeauthoryear{Tramper et al.}{2016}]{tramper2016} Tramper F., Sana H., Fitzsimons N. et al., 2016, MNRAS, 455, 1275

\bibitem[\protect\citeauthoryear{Tutukov \& Yungelson}{1973a}]{tutukov1973a} Tutukov A. V., Yungelson L. R., 1973, Nauchnye Informatsii, 27, 58

\bibitem[\protect\citeauthoryear{Tutukov \& Yungelson}{1973b}]{tutukov1973b} Tutukov A. V., Yungelson L. R., 1973, Nauchnye Informatsii,  27, 70

\bibitem[\protect\citeauthoryear{Tutukov \& Yungelson}{2002}]{tutukov2002} Tutukov A. V., Yungelson L. R., 2002, Astronomy Reports, 46, 667

\bibitem[\protect\citeauthoryear{van den Heuvel \& Heise}{1972}]{heuvel1972} van den Heuvel E. P. J., Heise J., 1972, Nature Physical Science, 239, 67

\bibitem[\protect\citeauthoryear{van den Heuvel \& De Loore}{1973}]{heuvel1973} van den Heuvel E. P. J., De Loore C., 1973, A\&A, 25, 387

\bibitem[\protect\citeauthoryear{van den Heuvel \& Yoon}{2007}]{heuvel2007} van den Heuvel E. P. J., Yoon S.-C., 2007, Ap\&SS, 311, 177

\bibitem[\protect\citeauthoryear{van den Heuvel, Portegies Zwart \& de Mink}{2017}]{heuvel2017} van den Heuvel E. P. J., Portegies Zwart S. F., de Mink S. E., 2017, eprint arXiv:1701.02355

\bibitem[\protect\citeauthoryear{van der Hucht}{2001}]{hucht2001} van der Hucht K. A., 2001, New Astronomy Reviews, 45, 135

\bibitem[\protect\citeauthoryear{Woosley}{1993}]{woosley1993} Woosley S. E., 1993, ApJ, 405, 273

\bibitem[\protect\citeauthoryear{Woosley}{2016}]{woosley2016} Woosley S. E., 2016, ApJL, 824, L10

\bibitem[\protect\citeauthoryear{Yoshida et al.}{2016}]{yoshida2016} Yoshida T., Umeda H., Maeda K., Ishii T., 2016, MNRAS, 457, 351

\bibitem[\protect\citeauthoryear{Zdziarski et al.}{2013}]{zdziarski2013} Zdziarski A. A., Mikolajewska J., Belczynski K., 2013, MNRAS, L104

\end{thebibliography}
\end{document}